# Synthesis and Structural Analysis of an Emissive Colloidal Argyrodite Nanocrystal: Canfieldite Ag$_8$SnS$_6$

Francisco Yarur Villanueva,[1,2] Victor Quezada Novoa,[3] Pascal Rusch,[1] Stefano Toso,[1] Maxwell W. Terban,[4] Yurii P. Ivanov,[1] Joaquin Carlos Chu,[2] Maxine J. Kirshenbaum,[2] Ehsan Nikbin,[5] Maria J. Gendron Romero,[2] Mirko Prato,[1] Giorgio Divitini,[1] Jane Y. Howe,[5] Mark W.B. Wilson,[2*] Liberato Manna[1*]

[1]Istituto Italiano di Tecnologia, Via Morego 30, 16163, Genova, Italy

[2]Department of Chemistry, University of Toronto, Toronto, ON, M5S 3H6, Canada

[3]Department of Chemistry and Biochemistry and Centre for NanoScience Research, Concordia University, 7141 Sherbrooke Street West, Montréal, Quebec H4B 1R6, Canada

[4]Momentum Transfer GmbH, Luruper Hauptstraße 1, 22547 Hamburg, Germany

[5]University of Toronto, Department of Materials Science and Engineering, Toronto, ON, M5S3E4, Canada

**ABSTRACT:** We resolve a phase identification controversy in the Ag-Sn-S material system by unraveling the polymorphic structure of nanocrystals within the argyrodite material family. Argyrodites are a class of superionic materials used in their bulk form for applications in solid-state batteries and thermoelectrics, where their advantageous properties relate to their polymorphism. However, despite their well-studied bulk applications, the limited exploration at the nanoscale has left considerable potential for the discovery of emerging properties due to size effects. Further, phase identification presents a prominent challenge to the study of polymorphs in superionic conductors and related materials. In this work, we synthesize canfieldite-like (Ag$_8$SnS$_6$) nanocrystals to understand their formation and structural behavior at the nanoscale. We observe the emergence of emissive, meta-stable, cluster-like species. Then, high-resolution transmission electron microscopy reveals indistinguishable polymorphs of canfieldite due to identical heavy-atom frameworks. However, using synchrotron X-ray total scattering for pair distribution function analysis, we uncover structural distortions, showing a pseudo-orthorhombic configuration that likely gives rise to the red emission. Further, we investigate the optical properties and structure of Ag$_8$SnS$_6$ nanocrystals upon the addition of Zn$^{2+}$, the cation of interest in the canfieldite vs. pirquitasite (Ag$_2$ZnSnS$_4$) phase identification controversy. We show that Zn$^{2+}$ is incorporated in the canfieldite-like structure through the replacement of Ag$^+$, boosting the emission. Our results solve a standing phase identification challenge and uncover fundamental insights for the synthesis and structure of canfieldite nanocrystals, laying the ground for the exploration of other argyrodite materials with emerging properties at the nanoscale.

## INTRODUCTION

Semiconductor nanocrystals (NCs) are technologically relevant materials due to their size-tunable optoelectronic properties.[1-2] Historically, lead, cadmium, and mercury-based binary systems have been at the research forefront.[3] However, their high environmental toxicity has led to restrictions by the European Union.[4] Therefore, there is a need for less-toxic NC alternatives. In this regard, the list of non-restricted materials for optoelectronic devices is limited. For instance, InP is the principal alternative to replace Cd-containing NCs.[5] However, the scarcity of indium and its environmental concerns highlight the need for other options.[6] Similar concerns apply to many emissive NCs operating in the visible and near-IR region, leaving an opportunity for more sustainable solutions. A viable strategy to find Pb/Cd-free alternatives that perform as well as their restricted counterparts is to increase the complexity of NCs by exploring multinary compositions, which would drastically increase the number of candidate materials compared to binary materials alone.[6]

In this regard, Ag-containing NC alternatives have garnered attention in the past years due to their low toxicity and large variety of phases.[7-9] However, only a limited number of Ag-based systems has been explored through colloidal chemistry and further advancements have been hampered by challenges including limited thermal conductivity, poor phase stability, and low photoluminescence quantum yields.[10-11] An attractive family in the compositional space of Ag-based ternary materials is the argyrodites family, which share a common formula of $A^{m+}_{(12-n)/m}B^{n+}X^{2-}_6$, (where A = Li, Ag, Cu; B = Si, Ge, Sn; and X = S, Se, Te with m and n as valence states of A and B, respectively).[12] Bulk argyrodites have complex and flexible lattices: cation disorder, for instance, is a major feature of these lattices because of their weak metal-to-chalcogen bonding.[12-15] Still, controlling cation disorder and ionic conductivity renders materials useful in band gap tunability,[16] photovoltaics and battery materials,[8,17-18] and thermoelectrics,[19] making argyrodites highly appealing materials.

Among the naturally occurring argyrodites, canfieldite (Ag$_8$SnS$_6$) is a highly attractive optoelectronic material due to its direct and narrow bandgap (1.1-1.4 eV) and high absorption coefficients (10$^4$-10$^5$ cm$^{-1}$).[20-23] Indeed, some preliminary synthetic routes exist to obtain Ag$_8$SnS$_6$ as colloidal NCs and inks (ranging from 7 to 13 nm),[22, 24] one in which the modification of optical properties *via* quantum confinement has been suggested.[23] However, a significant challenge concerns the identification of argyrodite phases at the nanoscale (< 10 nm). This stems from the tendency of Ag-Sn-S and related systems to produce multiple phases with similar compositions and polymorphs which, combined with broadened X-ray diffraction (XRD) patterns due to the finite size of NCs, complicates structure identification.[23, 25-26] In particular, bulk Ag$_8$SnS$_6$ is known to crystallize into two polymorphs, the room-temperature orthorhombic (Pna2$_1$) and the high-temperature cubic ($F\bar{4}3m$) one above ~170 °C. This phase transition is attributed to the onset of superionic conductivity, which arises due to the thermal disordering of the cationic positions of Ag$^+$ in the crystal.[14, 21, 27-28]

Polymorph identification and functionality are critical considerations in the development of multinary NCs.[13, 29-31] For instance, the presence of polymorphs in systems such as Cu$_2$ZnSnS$_4$ and AgBiS$_2$ can profoundly impact device performance.[32-34] Additionally, a significant challenge has been found in the phase identification of Ag-Sn-S systems with added zinc (*e.g.*, Ag$_8$SnS$_6$ *vs.* Ag$_2$ZnSnS$_4$). This difficulty arises from the ambiguous experimental stoichiometries and diffractograms observed,[7, 35-37] complicating structural characterization and posing the question as to whether Zn$^{2+}$ is incorporated into Ag$_8$SnS$_6$ particles to yield Ag$_2$SnZnS$_4$ in any measurable quantity. Considering the polymorphic nature of argyrodites and the expected roles of cationic mobility and structural distortion,[3, 22] clarification of these points is of central importance to establish design principles in this family of low elemental toxicity NCs with useful optoelectronic properties.

Here, we investigate the growth and controversial structure of NCs in the extended argyrodite family that contains Ag, Sn, Zn, and S as an attractive, Pb/Cd-free material system. We synthesized sub 7 nm Ag$_8$SnS$_6$ NCs, with emission in the λ: 750-830 nm region. Under our reaction conditions, we can also observe the formation of an emissive, cluster-like species (~1.5 nm in diameter) with peak emission at λ: 630 nm, which is the first observation of such a species to the best of our knowledge. Then, we resolve the phase identification challenge in Ag-Sn-S systems by employing elemental analysis techniques as well as high-resolution (scanning) transmission electron microscopy (HR-STEM) and synchrotron X-ray total scattering for pair distribution function analysis (PDF). Our findings demonstrate that distinguishing between the orthorhombic and the cubic polymorphs of canfieldite becomes unimportant at the nanoscale because NCs adopt a canfieldite-like phase with a pseudo-orthorhombic structure of Ag$_8$SnS$_6$. Finally, we exploited the same set of techniques to retroactively investigate canfieldite NCs prepared in the presence of Zn$^{2+}$, which were previously believed to be a different material (*i.e.*, pirquitasite Ag$_2$ZnSnS$_4$). Contrary to expectations, we demonstrate that the incorporation of Zn$^{2+}$ does not change the canfieldite-like phase but it contributes to boosting the emission. Overall, this investigation into the synthesis, optical properties, and structural composition of Ag-Sn-S NCs expands our understanding of superionic semiconductor materials at the nanoscale, opening up new possibilities for material design within the argyrodite family.

## RESULTS AND DISCUSSION
### 1. Optical and structural properties of NCs in the Ag-Sn-S (ATS) material family
*1.1. Synthesis and Optical Properties of ATS Products*

We first focused on the synthesis and characterization of the optical properties of ATS NCs with the goal of learning about this underexplored system as well as having optical information that could be correlated to our structural findings. To synthesize ATS NCs, we modified our reported procedure to target pure ATS NCs (See Methods).[35] Our ATS syntheses yield products with sizes 2.1-6.9 nm under low-resolution TEM (Figure S1). The size can be tuned by varying the growth time and reaction temperature while the isolation requires a size-selective precipitation (See Methods and Figure S2). Visually, solutions with smaller ATS NCs have a bright red colour while the fraction of larger NCs has a dark brown tint, while optical spectroscopy reveals corresponding shifts in their absorption spectra emission peak (Figure 1a). Notably, the peak emission at of the smaller species is located with remarkable consistency at λ: 630 nm across syntheses (Figure 1a & S3). By comparison, the emission peaks from larger ATS NCs are in the λ: 700-750 nm range (Figure 1 a), consistent with a size effect, but vary between batches and fractions (Figure S3). Further, samples of the smaller red-looking species are significantly unstable and convert to the brown species with emission near λ: 740 nm (typical of larger ATS NCs) over a ~12 hr period (Figure S4). These observations align with the behaviour of thermodynamically-unstable cluster species seen in other material systems, suggesting that such species might be present in our synthesis.[38-40]

Elemental analysis through STEM-EDX and XPS of ATS NCs with an average size of 6.0 nm show a stoichiometry of 4.0:1.0:3.2 and 3.3:1.0:3.6 (Ag:Sn:S), respectively (Figure S5, S6, and Table S1). These results are *not* consistent with the theoretical stoichiometry of bulk canfieldite Ag$_8$SnS$_6$ and could align with other phases in the Ag-Sn-S system such as monoclinic Ag$_2$SnS$_3$ or Ag$_4$Sn$_3$S$_8$ (Figure S7c). However, major variations in the theoretical stoichiometry can be observed in NCs due to faceting, where the atoms missing from the truncated lattice can be strongly biased by facet-specific surface energies and ligand interactions.[41-43] Thus, an orthorhombic canfieldite structure for our ATS NC is similarly plausible and the structure cannot be attributed solely through compositional analysis.

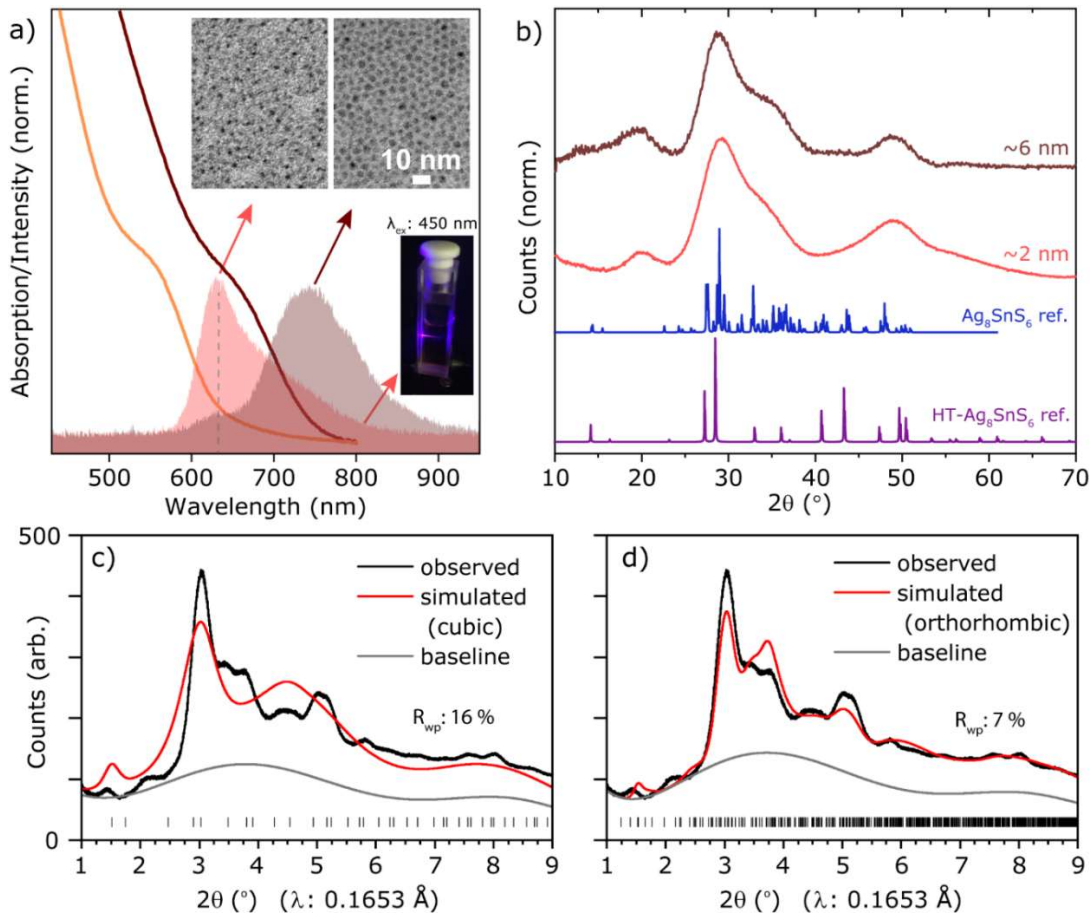

**Figure 1**. Optical characterization, TEM images, and diffractograms for ATS NCs. a) Absorbance and photoluminescence spectra for ATS NCs purified through a size-selective precipitation. The cluster-like species have a peak emission at 630 nm. The inset shows bright-field TEM images (120 kV) of these two NCs samples. The bottom inset shows a cuvette containing cluster-like ATS NCs being illuminated by λ: 450 nm light. b) Experimental PXRD diffractograms for ATS NC samples of sizes 2 and 6 nm, acquired for 24 and 6 hours, respectively. The reference patterns for orthorhombic (PDF 00-038-0434) and high-temperature cubic (PDF 04-002-4840) canfieldite are shown in blue and purple. Rietveld fit for patterns acquired on 6 nm NCs for the cubic c) and orthorhombic d) phase using identical baseline functions. $R_{wp}$ represents the weighted profile residual.

PXRD measurements on ~ 2 and ~ 6 nm NCs are ambiguous and do not allow us to confidently assign a phase, showing significantly broadened Bragg reflections (Figure 1 b). Indeed, the experimental patterns look unusually similar considering that one would expect a 3-fold sharpening in the signal as per the Scherrer equation, assuming single-domain NCs (Figure S7a, *vide infra*).[44] We then acquired high-resolution PXRD data to test the cubic and orthorhombic models, and performed Rietveld refinement on the resulting diffractograms (See SI Section 2 for details). Our analysis shows that the cubic structure cannot index the second feature just above 2 ° and fits poorly to the rest of the reflections (Figure 1c). On the other hand, the orthorhombic structure better describes the features of the diffraction pattern but also does not give a good match (Figure 1d). Overall, due to extreme broadening of Bragg reflections and limited description of the diffraction features by either model, neither can accurately describe the structure of the particles, and further analysis of the local structure is required, vide infra.

*1.2. Single-NC Analysis via HR-TEM*

In order to simplify our structural analysis and facilitate comparison to literature data, we decided to focus on larger (>5 nm, Figure 2a) ATS NCs, which are grown at 95 °C (See Methods). Since structural analysis using PXRD yields ambiguous results regarding the phase, we acquired single-NC images by means of HR-TEM to probe the crystalline structure of ATS NCs (Figure 1b). In the TEM images we note ATS NCs that display signs of polycrystallinity (Figure 2b and S8). We identify NC structures that appear to be composed of smaller triangular crystalline domains of height ~2.1 nm (Figure 2c). Such triangular domains are in the length order of 2-3 canfieldite unit cells and aggregate in an ordered fashion, resembling the behaviour of some gold nanoclusters, which undergo aggregative growth to form star-shaped particles.[45] Furthermore, we noticed that these polycrystalline NCs coalesce under electron beam irradiation into single, but faulted, crystalline domains (Figure S8), aided by the high intrinsic mobility of Ag atoms in argyrodite materials.[12-13, 15] These results provide insight into the formation mechanism of ATS NCs in which a stepwise process concerning the pre-formation of cluster-like species followed by aggregative growth and coalescence could be at play. In fact, this hypothesis can explain the optical behaviour that we see in both the absorption and PL spectra in Figure 1a where the emission of cluster-like species at 630 nm and the excitonic features broaden as NCs grow. Additionally, polycrystallinity and faulting in NCs have been correlated with electronic trap states, quenching the PL.[46-47] This also aligns with the low PLQY value (<1 %) that we measure for these particles. The bottom scheme in Figure 2 summarizes our proposed formation mechanism and aggregative growth step for ATS NCs. Structurally, such degree of polycrystallinity and stacking faults are known to affect NC phase identification, provoking significant broadening and changes to PXRD diffractograms.[48-50] Thus, our observations provide a partial explanation of the breadth and unexpected size-independent of the Bragg reflections in our PXRD measurements (Figure 1b and S7a). However, bypassing polycrystallinity likely requires reaction temperatures exceeding 400 °C as suggested on our temperature-dependent PXRD measurements (Figure S9). While this falls outside the scope of our current study, optimizing crystallinity under milder conditions presents an important avenue for future research.

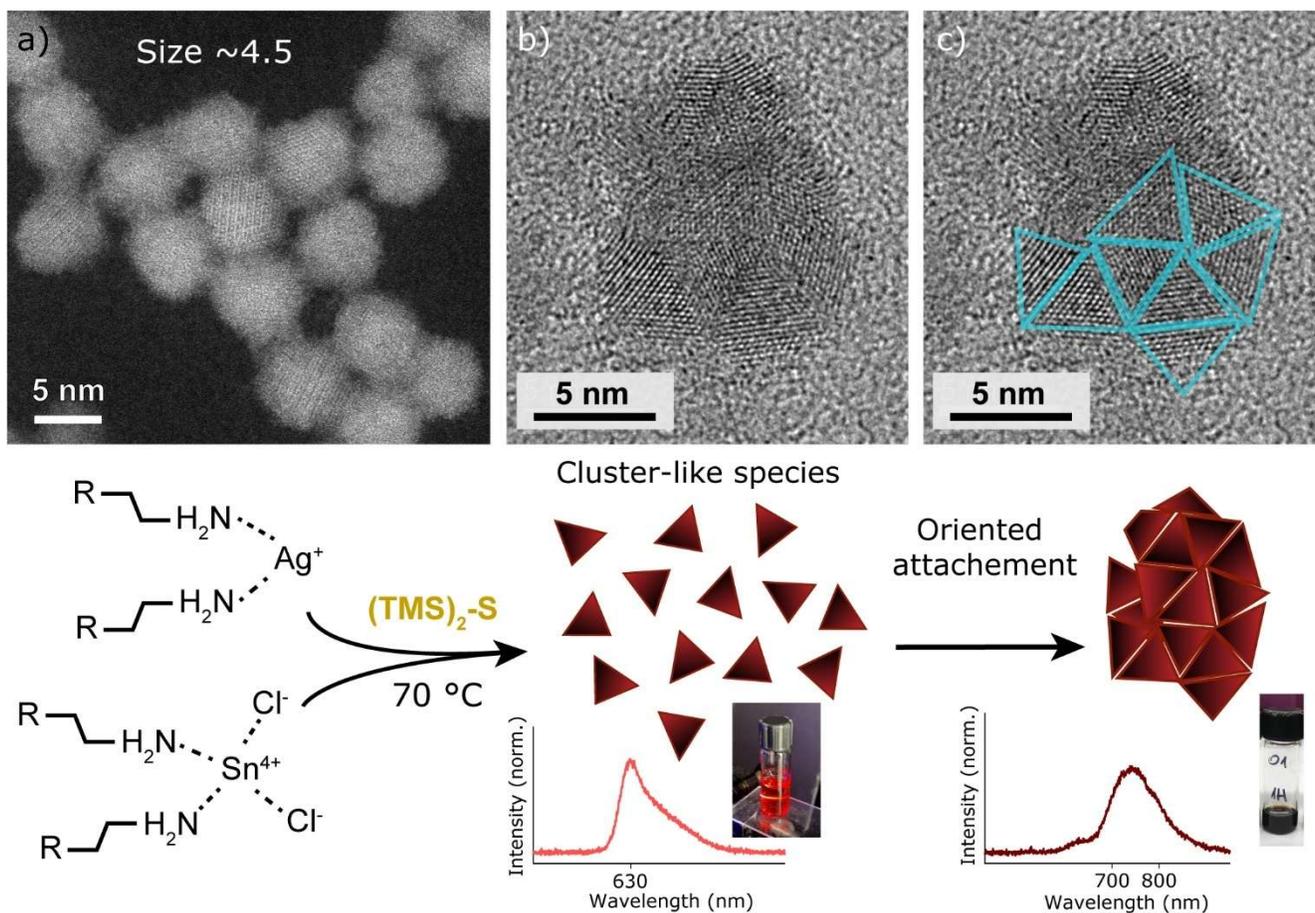

**Figure 2.** TEM images and reaction scheme for ATS NCs. a) HR HAADF STEM image showing ATS NCs of size ~4.5 nm. b and c) Bright field HR-TEM image illustrating a polycrystalline ATS NCs composed of cluster-like species. The cartoon at the bottom depicts the proposed formation mechanism of cluster-like species and their aggregative growth leading to larger, polycrystalline ATS NCs

We then performed FFT analysis to probe the phase of ATS NCs. Through our measurements, we identify images displaying the [211] and the [110] zone axes of canfieldite with d-spacing 0.635, 0.388, and 0.558 nm (Figure 3a–e). These d-spacing values have an excellent correlation with reported values for the high-temperature cubic phase of canfieldite where silver is highly disordered.[14] Additionally, we identify d-spacings 0.311 and 0.301 nm that correlate with main reflections (022) and (411) of the orthorhombic phase of canfieldite (Figure S10).[51] However, despite the distinction in space groups, both phases share a common Sn framework (Figure 3e) with the primary difference being the degree of disorder in the silver atoms. As a result, these measurements robustly confirm that the position of heavy Sn atoms in these NCs is compatible with canfieldite. These findings also corroborate that the coalescence of polycrystalline NCs into single crystalline domains does not change the material composition. However, the weaker and broader reflections anticipated from the disordered Ag atoms mean that unraveling the structure of ATS NCs remains challenging *via* HR-TEM measurements alone. Therefore, we pursued further structural studies using pair distribution functional analysis because this technique is sensitive to the local coordination environment of atoms in the lattice.

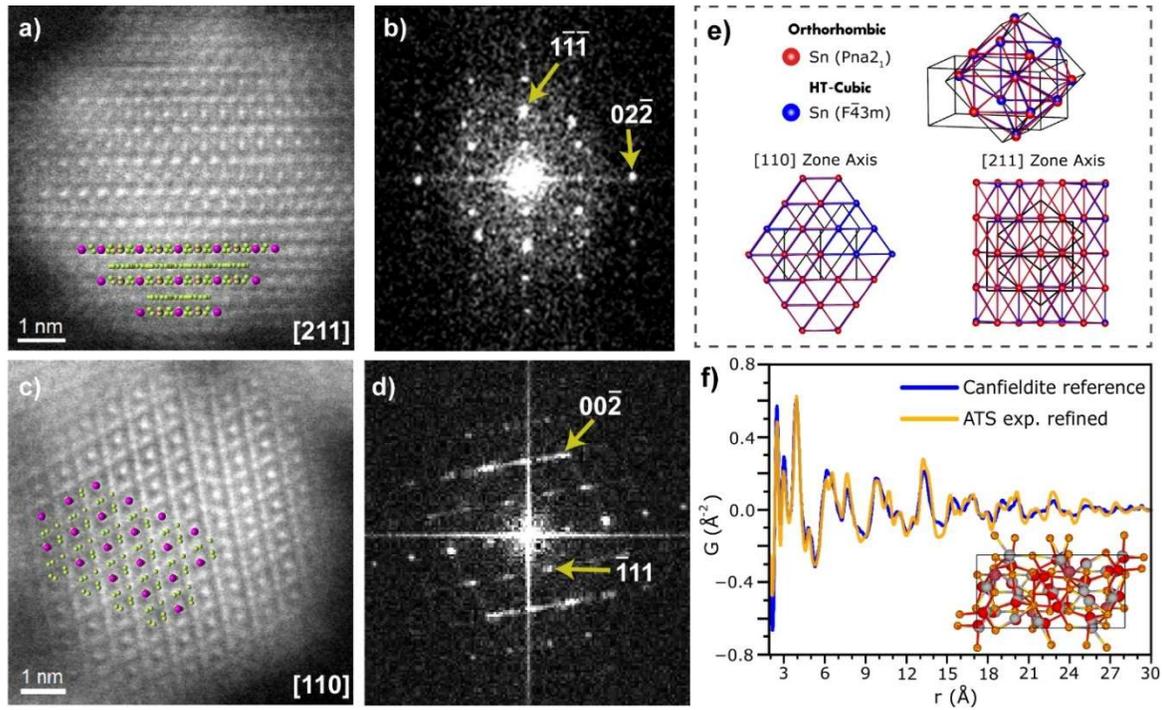

**Figure 3.** HR STEM images, corresponding FFT patterns, and PDF scattering pattern for ATS NCs. a) HR-HAADF-STEM image displaying the [211] zone axis of a canfieldite NC with the atomic map overlaid on the image for better visualization of the lattice with Sn in pink and Ag in green, b) associated FFT pattern. c) HR-HAADF-STEM image displaying the [110] zone axis for a canfieldite NC, d) associated FFT pattern. e) Atomic structural models showing the framework of Sn atoms in orthorhombic (red, PDF 00-038-0434) and cubic (blue, PDF 04-002-4840) canfieldite overlayed. Sn-Sn bonds do not exist in the original lattice but were simulated for better visualization of the indistinguishable frameworks. f) PDF analysis of ATS NCs. The Canfieldite reference model is fitted to the experimental data with refined Ag positions. The inset shows the structural model with refined Ag positions in red compared to the original atoms in grey.

### 1.3. Structural Analysis of ATS NCs via Pair Distribution Function.

We performed synchrotron X-ray total scattering measurements with PDF analysis on ~6 nm ATS NCs to get further information about the structure of our NCs. The experimental PDF displays a loss in spatial coherence at very low distances (~30 Å), as seen by the dampening of the peak intensity (Figure 3f). This evidence suggests that the ordered domain size of our ~6 nm ATS NCs is of 3 nm, correlating with our observations of polycrystalline NCs (Figure 2b, c) and providing further support for our proposed growth model via aggregative growth (scheme in Figure 2). To investigate the structure of our particles, the PDF data were initially fitted using the model of bulk orthorhombic canfieldite (Figure 3f). Our model was refined by adjusting the position of all Ag atoms while keeping both the unit cell parameters as well as the position of Sn and S atoms identical to the original orthorhombic model. This refinement described the local structure (2-10 Å) of ATS NCs with a goodness-of-fit ($R_w$) of 0.257, representing the main features of the experimental PDF without major differences from the original canfieldite model. This suggests that the local coordination environment of our ATS NCs is similar to canfieldite. However, considering the similarity in the Sn position between the orthorhombic and cubic structures (Figure 3e), and the highly disordered position of Ag atoms in the high-temperature cubic model, a more intricate modeling strategy was implemented to try to better capture and describe the local structure of ATS NCs. First, both orthorhombic and cubic models were evaluated in terms of their capacity to describe the intermediate-to-long-range structure (10-30 Å), which is subject to long-range averaging of different local environments in a statistical structure.[52] In this analysis, both models described the overall features similarly (Figure S11): cubic model with disordered Ag sites ($R_{wp}$ = 28 %), orthorhombic model with discrete Ag sites ($R_{wp}$ = 29 %). Thus it is possible that the description of the local structure could be better explained by a model that contains disordered Ag atom sites. To evaluate this hypothesis, we constructed four different models comprised of the following basis features: $SnS_4$ tetrahedra modeled as rigid bodies. The Sn atom was fixed while the vertices of the tetrahedra (S atoms) were allowed to rotate around the center of mass, and Ag atoms with positions refined freely except for an anti-bump constraint (more details about model characteristics are found in SI Section 4 and Table S2). The four models are referred to as pseudo-cubic, orthorhombic, pseudo-orthorhombic I, and pseudo-orthorhombic II. In our analysis, we observed that the system was insensitive to the rotation of SnS$_4$ tetrahedra due to their low contribution to the PDF signal. Thus, this feature was set to fixed positions in subsequent calculations, a reasonable assumption, considering that the framework of Sn atoms is identical in both polymorphs (Figure 3e). Furthermore, to account for Ag atom disorder, our models required discrete Ag positions.

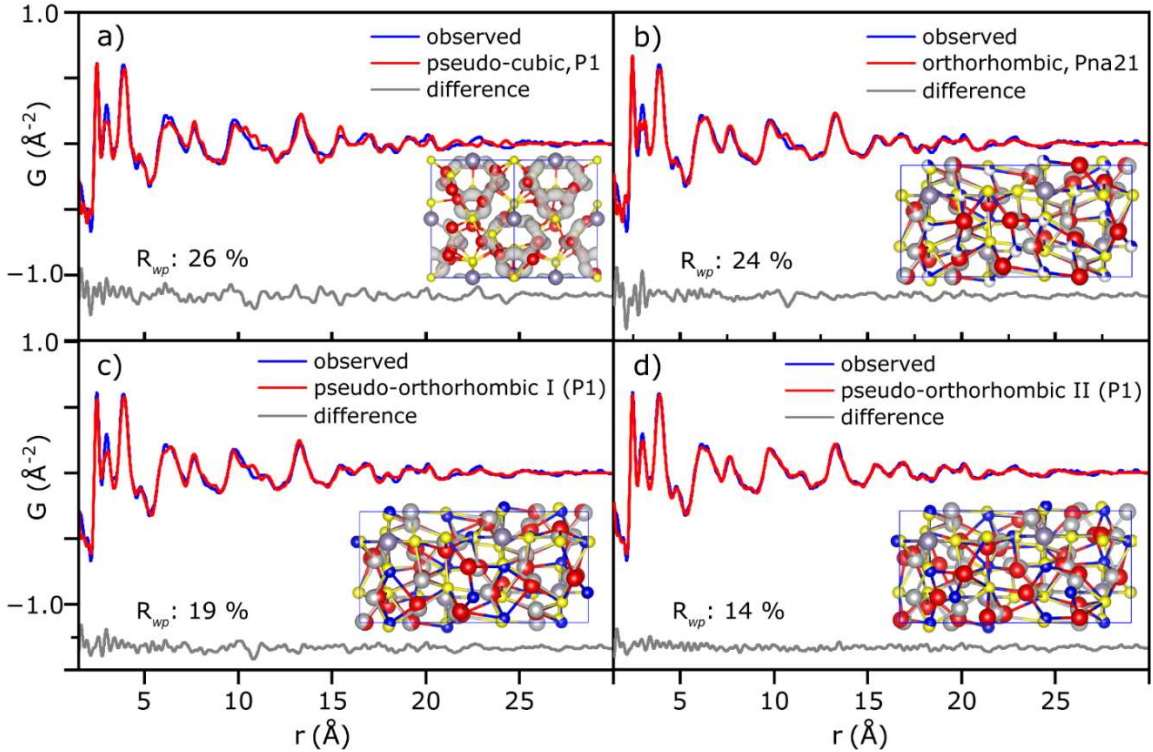

**Figure 4**. Results of real-space fits and models of the modified cubic and orthorhombic structures to the local and intermediate range structuring as observed in the PDF for a) pseudo-cubic, b) orthorhombic, c) pseudo-orthorhombic I, and d) pseudo-orthorhombic II. Each inset structure is an overlap of the original position of Ag (grey), Sn (purple), and S (yellow) atoms and the refined positions (red). $R_{wp}$ represents the weighted profile residual. In a) the isosurface describes the disorder of Ag atoms in the original cubic model.

Considering these two requirements, the pseudo-cubic (Figure 4a) and orthorhombic (Figure 4b) models underperform in the fit compared to the pseudo-orthorhombic models (Figure 4c and 4d). This suggests that Ag prefers an arrangement of SnS$_4$ tetrahedra that are slightly distorted away from their relative positions compared to the cubic model, making it difficult to clearly distinguish the structural tiling. Such distortion is feasible if we consider that the distribution of local environments is more diverse than what can be captured in this simple small-box model. The pseudo-orthorhombic II (P1) structure was used for Rietveld analysis of the ATS diffraction pattern (Figure S12) showing the capabilities of the model to describe the broad reflection features of the sample, as described by preliminary Rietveld analysis shown above.

Taken together, our models represent a description and compatibility of atomic positions with the local structure of ATS NCs, which in reality is a much broader distribution of local structures. Ultimately, PDF analysis suggests that: 1) the local and intermediate-range structure of ATS NCs is compatible with a canfieldite-like structure, 2) the true structure likely comprises site disorder of the Ag atoms, distributed along a tiling of SnS$_4$ tetrahedra and S atoms, and 3) SnS$_4$ tetrahedra can rotate on site in the structure, however its PDF signal does not appear sensitive enough to finely resolve the nature of their orientations.

*1.4. Structural Analysis of Cluster-Like Species*

We performed total X-ray scattering measurements and PDF analysis on ATS cluster-like species to corroborate the structure and get additional insight into their size. Given that these species would degrade rapidly (Figure S4), we sought a post-synthetic stabilization procedure using ZnBr$_2$ to extend their bench lifetime despite their minimal compositional effects, which do not affect the structural symmetry (See Section 2). The PDF data matched that of the two pseudo-orthorhombic models at short distances (Figure S13). However, a major difference between large ATS NCs and these cluster-like samples is the rapid and sharp signal dampening at ~5 Å and loss of spatial coherence at ~15 Å. This value is on the order of a few canfieldite unit cells (a: 15.3, b: 7.5, c: 10.7 Å). Hence, this result supports our hypothesized formation of a small, cluster-like species with canfieldite structure. Additionally, Figure S12c shows that the comparison between the experimental diffraction pattern of cluster-like species and that of the pseudo-orthorhombic (P1) structure model after refinement through PDF analysis. Their resemblance demonstrates the ability of our model to describe the structure of cluster-like species. Overall, the PDF analysis of our ATS particles reveals clear structural deviations compared to bulk canfieldite, suggesting that Ag and Sn atoms play an important role in the optical properties that emerge at the nanoscale (*i.e.,* red emission). Furthermore, our PDF analysis displays evidence for the presence of canfieldite cluster-like species, which expands our understanding of the formation mechanism of larger NCs in this material family.

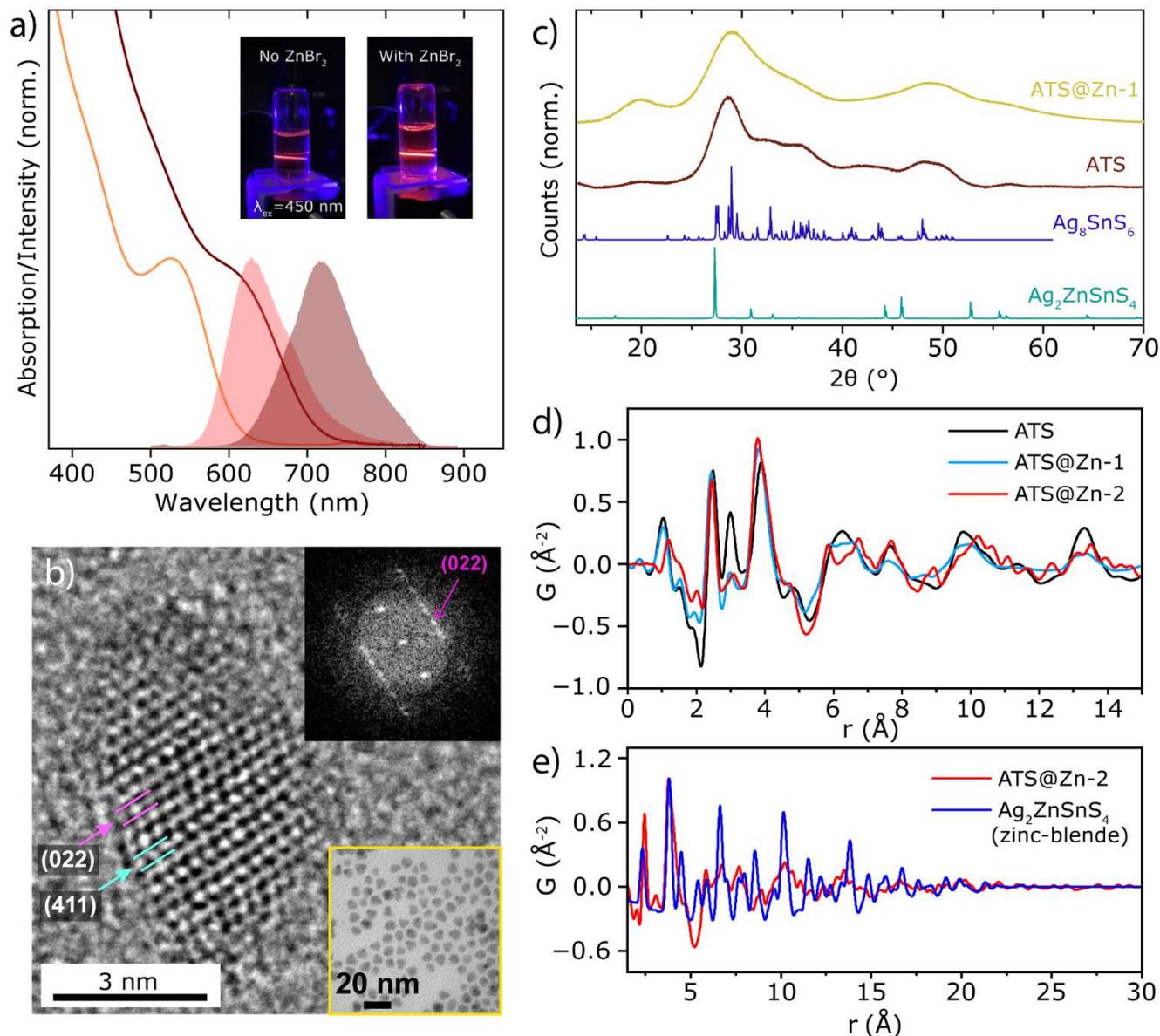

**Figure 5**. Optical and physical characterization for ATS NCs containing Zn. a) Optical characterization for two ATS@Zn-1 NC sizes. The absorbance spectrum of the smallest NCs (orange trace) shows a well-defined excitonic feature and an emission peak at ~630 nm seen in the red trace and the inset. The brown traces display the absorbance and photoluminescence spectra for larger ATS@Zn-1 NCs. b) HR-PXRD data for ATS and ATS@Zn-1. c) HR-TEM image and corresponding FFT pattern for a single ATS@Zn-1 NCs, displaying d-spacing values that correlate with canfieldite. The inset shows a low-resolution TEM image of ATS@Zn-1 NCs of size ~6 nm used for PDF analysis. d) PDF scattering plots for ATS, ATS@Zn-1, and ATS@Zn-2. e) PDF scattering plot comparing ATS@Zn-2 to bulk $Ag_2ZnSnS_4$. The local structure of ATS@Zn-2 NCs is not compatible with the zinc-blende phase

## 2. Effect of Zn on the structural and optical properties of ATS NCs. Solving the phase identification controversy.

After unraveling the structural characteristics of pure ATS NCs, we proceeded to tackle the structural controversy into the phase identification of Ag-Sn-S systems with added $Zn^{2+}$ (which we will refer to as ATS@Zn). More specifically, our motivation stems from the fact that our groups and others[7, 35, 37] have encountered phase identification challenges when studying the synthesis and physicochemical properties of supposed pirquitasite ($Ag_2ZnSnS_4$) NCs. Similar to the discussion of canfieldite NCs above, the challenges arose in part due to the severe broadening of Bragg reflections.[7, 35, 53-54]

### 2.1. Synthesis and Optical Properties of ATS@Zn Products

We introduced $Zn^{2+}$ into ATS reactions (See Methods), following the methods previously reported to yield pirquitasite NCs. ATS@Zn samples were prepared using our reported procedure[7, 35] as well as a previously developed synthesis[7, 35] (See Methods). We will refer to these NCs as ATS@Zn-1 and ATS@Zn-2, respectively. These samples were of similar average size (~5.0 nm, under low-res TEM) with comparable stoichiometry to ATS NCs and had a photoluminescence maximum at ~ 790 nm (Figure S14 and Table S1). Visually, we can see that the presence of $Zn^{2+}$ grants a more-gradual NC growth in ATS@Zn-1 products. We speculate that $Zn^{2+}$ regulates the kinetics by acting as a Z-type ligand on surface sulfur, as seen in other materials.[55-56] $Zn^{2+}$ may also slow growth by partially exchanging with $Ag^+$, reducing free $Ag^+$ availability for lattice incorporation. Further, the excitonic absorption feature is more defined in ATS@Zn-1 (Figure 5a) relative to ATS NCs (Figure 1a). This allows us to better follow the growth (inferred from a progressive bathochromic shift) of these species (Figure S15). The peak of the excitonic absorption in the earliest aliquots is at 2.36 eV (Figure 5a), representing an optical gap that is 1.11 eV larger than the bandgap of bulk canfieldite. We also observe a PLQY enhancement from ~1 to 5% when ATS NCs are treated with $ZnBr_2$ post-synthetically, matching previous reports and suggesting that in our case, Zn may also passivates surface traps.[57-59] In addition, elemental analysis using XPS shows average elemental ratios of 2.2:1.0:0.7:3.9 for ATS@Zn-1 NCs (Ag:Sn:Zn:S, Figure S5 and Table S1). This observation is consistent with previous comparable studies,[7, 35, 37] wherein it was used to support claims of the synthesis of pirquitasite ($Ag_2SnZnS_4$) as the primary phase. This is because the roughly 1:1 ratio between Sn and Zn is consistent with the bulk stoichiometry of pirquitasite. However, considering that the unit cell of canfieldite contains only four Sn atoms, and that larger NCs are composed of aggregates of smaller domains (*vide supra*), we speculate that even surface-bound Zn could contribute significantly to the stoichiometry of the overall NC. Any *de facto* replacement of Ag by Zn could also cause stoichiometric deviations. Therefore, we explored whether a canfieldite core passivated with Zn on the surface could also explain the composition that we observe.

2.2. *Single-NC Analysis via HR-TEM*

We conducted HR-TEM on ATS@Zn-1 NCs to obtain structural information about the impact of Zn passivation and/or incorporation. The images show similar results to ATS NCs in terms of d-spacing values where we identify the (411) and (022) planes of canfieldite (Figure 5b). Additionally, we attain further insight into the preferred planes for the formation of stacking faults (See Figure S16 and associated discussion). However, it is important to highlight that relying solely on d-spacing values can lead to ambiguity over the phase distinction between pirquitasite and canfieldite. This is because the (112) and (022) lattice planes (and main PXRD reflections) in pirquitasite and canfieldite respectively have the same interplanar distance (*i.e.,* 0.31 nm). Consequently, based on our HR-TEM measurements we cannot confirm the presence of pirquitasite or the inclusion of $Zn^{2+}$ in the lattice in ATS@Zn-1 products. Instead, our results have a better correlation with the structural characteristics of a canfieldite-like phase.

2.3. *Crystallographic Analysis for ATS@Zn NCs via Pair Distribution Function*

We used synchrotron total X-ray scattering measurements on ATS@Zn to investigate how $Zn^{2+}$ affected the local structure of ATS and assess whether the formation of pirquitasite was feasible under our reaction conditions, as well as under previously reported conditions. Although the information extracted through high-resolution PXRD is limited due to the breath of the Bragg reflections, comparing the diffraction pattern of ATS to that of ATS@Zn-1 NCs provides qualitative insight. Specifically, all the reflections seen for ATS@Zn-1 NCs are notably broader than those of ATS NCs (Figure 5c), which would arise if the addition of Zn resulted in smaller effective crystal domains. PDF analysis of ATS@Zn-1 NCs reveals a signal that resembles that of pseudo-orthorhombic canfieldite-like models (Figure 5d and S16). However, the nearest-neighbor distances for Sn-S and Ag-S pairs in ATS@Zn-1 (~2.421 Å) are shorter than those in pure ATS (~2.464 Å). Given that $Zn^{2+}$ has a smaller ionic radius than $Ag^+$, this contraction is consistent with the formation of Zn-S pairs. Additionally, the peak at ~ 3 Å, primarily attributed to Ag-Ag pairs in ATS, exhibits significantly reduced intensity. We attribute this weakening to the introduction of Zn-Ag pairs because Zn has a lower scattering power, diminishing the overall signal. Similarly, the peak at ~ 4 Å, dominated by Ag-Sn pairs, shifts to a shorter distance. We associate this with the incorporation of Zn-Sn pairs in ATS@Zn-1, leading to a reduced next-neighbor distance in Zn-(S)-Sn and Ag-(S)-Sn configurations. Furthermore, the higher radial distances in ATS@Zn-1 appear slightly broader, and the domain size is smaller (~ 2.5 nm) compared to ATS (Figure 5d and S17), indicating greater structural disorder. This disorder likely stems from a wider distribution of local environments induced by $Zn^{2+}$ incorporation throughout the whole ATS structure. Synthetically, this means that the inclusion of $Zn^{2+}$ through $Ag^+$ atom displacement must occur *after* the formation of the ATS core, given that $Zn^{2+}$ is introduced post-nucleation, rendering our NCs a highly appealing platform for cation exchange.

We then proceeded to analyze NCs from a second reported procedure to achieve pirquitasite[7] (labeled ATS@Zn-2 here). The NCs used were of similar size and emission to our ATS@Zn-1 samples (Figure S14c and d). Unexpectedly, the PDF of ATS@Zn-2 NCs coincide with a canfieldite-like phase and **not** zinc-blende pirquitasite (Figures 5d and 5e). In fact, the zinc-blende fit cannot even remotely describe neither the local structure nor larger distances in ATS@Zn-2 samples (Figure 5e), evidencing that this structure is highly inaccurate to characterize these NCs. Conversely, comparing the experimental data for ATS@Zn-2 to the canfieldite-like model shows that the local structure up to 5 Å is highly similar to ATS and ATS@Zn-1 with small Zn-related modifications. We also noticed slight deviations beyond the local structure (> 5 Å) in ATS@Zn-2 compared to ATS@Zn-1 (Figure S17). We attribute this small difference to the way in which $Zn^{2+}$ is included in the structure. In ATS@Zn-1 the $Zn^{2+}$ inserts into the lattice after the ATS core is formed at 70 °C, while in ATS@Zn-2 the $Zn^{2+}$ is present in the reaction before the sulfur injection at 160 °C. Overall, the main features of the PDF of ATS@Zn-2 are highly compatible with a canfieldite-like model. Hence, our results indicate that the formation of true pirquitasite NCs might have a higher thermodynamic barrier relative to canfieldite and, thereby, be more difficult to access with conventional synthetic routes.

CONCLUSION

We investigated the synthesis and structural characterization of canfieldite ($Ag_8SnS_6$, ATS), an under-explored, superionic material system, in its nanoscale form. First, we found emissive cluster-like ATS species that are involved in the formation and growth of larger NCs *via* aggregative growth and coalescence. The red emission is not observed in bulk canfieldite and is likely related to size reduction. These findings set a baseline for the design of procedures for the synthesis of superionic argyrodite NCs. We performed in-depth structural studies *via* HR-STEM to unravel the structural features of ATS NCs. We find that the breadth in reported PXRD diffractograms likely arises due to polycrystallinity. Additionally, we find that the structure of our NCs is compatible with a canfieldite lattice, and we show that the framework of heavy atoms (*i.e.,* Sn) is indistinguishable between canfieldite polymorphs, an overlooked fact that frustrated previous interpretations of the phase of Ag-Sn-S materials. We modelled canfieldite polymorphs through PDF analysis and found that the structure of our NCs cannot be described through a single bulk-derived model but instead requires an averaged canfieldite-like phase that accounts for structural modifications. Such structural distortions observed in nanocrystalline samples of ATS correlate with the onset of photon emission, highlighting a possible role of nanoscale structural perturbations in dictating optoelectronic properties. Our results emphasize the importance of probing the local structure of NCs to fully understand and tailor the functionality of nanomaterials. Future work incorporating molecular dynamics simulations could provide deeper mechanistic insights into the dynamic behavior and structural flexibility of these systems.

Finally, we solved a standing controversy regarding the phase identification in canfieldite-pirquitasite ($Ag_2ZnSnS_4$) Ag-Sn-S@Zn systems using a combination of HR-TEM and pair distribution function. Critically, we identify that HR- measurements are unsuitable to confirm the presence of pirquitasite because of the similarity in d-spacing value with canfieldite. We performed total scattering measurements with PDF analysis on various Ag-Sn-S-Zn samples and confirm the formation of canfieldite as the main phase in all synthesized products. We find that the addition of $Zn^{2+}$ into the canfieldite syntheses does not grant the formation of pirquitasite $Ag_2ZnSnS_4$, as previously thought, even when following reported higher-temperature procedures. Instead, we find that $Zn^{2+}$ replaces $Ag^+$ throughout the lattice, making our particles an appealing platform for cation exchange procedures.

Taken together, our study elucidates the design and exploration of argyrodite nanostructures, revealing critical synthetic and structural insights that advance the understanding of their unusual nanoscale behavior. Future work should employ *ab initio* molecular dynamics to resolve static versus dynamic disorder, refining knowledge of structural distortions governing optoelectronic properties. Additionally, NMR studies will help reveal surface faceting and key binding sites to enhance the functionality of this material system. These findings will inspire synthetic and computational efforts to uncover emerging properties in under-explored multinary systems, moving beyond binary materials, while discovering effective dopants and surface stabilization techniques will be key to enhancing their performance.

## ASSOCIATED CONTENT

**Supporting Information**. Section 1 contains the full description of the synthetic methods and characterization techniques. Also, there are two tutorial videos demonstrating the synthetic procedures for ATS and ATS@Zn-1 NCs. Section 2 contains additional physical and optical characterization images for our products as well as a detailed description of the Rietveld refinement. Section 3 provides supporting TEM and diffraction data. Section 4 contains details about the PDF analysis including the parameters used for the modelling of the structure refinements. The CIF files for our four refined structures are also attached. Section 5 contains supplemental physical and optical characterization for products synthesized with added zinc.

## AUTHOR INFORMATION

Francisco Yarur Villanueva: 0000-0003-1102-3382

Victor Quezada Novoa: 0000-0001-8360-9847

Pascal Rusch: 0000-0001-5088-4447

Stefano Toso: 0000-0002-1621-5888

Maxwell W. Terban: 0000-0002-7094-1266

Yurii P. Ivanov: 0000-0003-0271-5504

Joaquin Carlos Chu: 0009-0001-6969-1191

Maxine J. Kirshenbaum: 0000-0002-4890-1137

Ehsan Nikbin: 0009-0008-8726-3626

Maria J. Gendron Romero: 0009-0003-2199-5107

Mirko Prato: 0000-0002-2188-8059

Giorgio Divitini: 0000-0003-2775-610X

Jane Y. Howe: 0000-0001-9319-3988

### Corresponding Author

* Liberato Manna: 0000-0003-4386-7985
E-mail: liberato.manna@iit.it
* Mark W.B. Wilson: 0000-0002-1957-2979
E-mail: mark.w.b.wilson@utoronto.ca

### Author Contributions

The manuscript was written through contributions of all authors. All authors have given approval to the final version of the manuscript.

## ACKNOWLEDGMENT

F.Y.V., J.C.C., M.K., M.J.G. and M.W.B.W. acknowledge the support of the Natural Sciences and Engineering Research Council of Canada (NSERC) via RGPIN2023-05041, as well as support for research infrastructure from the Canada Foundation for Innovation [JELF-35991], and the Ontario Research Fund [SIA-35991]. F.Y.V. and M.K. acknowledge support from NSERC Canada Graduate Scholarship–Doctoral (CGS-D) Fellowships, and F.Y.V. further acknowledges a Walter C. Sumner Memorial Fellowship, and an Irene R. Miller Scholarship in Chemistry. E.N. and Y.H. acknowledge Open Center for Characterization of Advanced Materials (OCCAM) for TEM support. Electron microscopy analysis was performed at the Italian Institute of Technology (IIT) with support from Luca Leoncino. XRD measurements were performed at the IIT with the assistance of Sergio Marras. P.R., S.T. and L.M. acknowledge funding from the Project IEMAP (Italian Energy Materials Acceleration Platform) within the Italian Research Program ENEA-MASE (Ministero dell'Ambiente e della Sicurezza Energetica) 2021-2024 "Mission Innovation" (agreement 21A033302 GU n. 133/5-6-2021). We acknowledge the European Synchrotron Radiation Facility (ESRF) for provision of synchrotron radiation facilities. We would like to thank the Momentum Transfer team for facilitating the measurements and Jakub Drnec for assistance and support in using beamline ID31. The measurement setup was developed with funding from the European Union's Horizon 2020 research and innovation program under the STREAMLINE project (grant agreement ID 870313). M.J.K. acknowledges the Brookhaven National Lab National Synchrotron Light Source II, beamline 28-ID-2 (proposal number: 315071) for facilitating synchrotron measurements under the assistance of Sanjit Ghose.

REFERENCES

1. Kagan, C. R.; Lifshitz, E.; Sargent, E. H.; Talapin, D. V., Building Devices from Colloidal Quantum Dots. *Science* **2016**, *353* (6302), aac5523.
2. Efros, A. L.; Brus, L. E., Nanocrystal Quantum Dots: From Discovery to Modern Development. *ACS Nano* **2021**, *15* (4), 6192-6210.
3. García de Arquer, F. P.; Talapin, D. V.; Klimov, V. I.; Arakawa, Y.; Bayer, M.; Sargent, E. H., Semiconductor Quantum Dots: Technological Progress and Future Challenges. *Science* **2021**, *373* (6555), eaaz8541.
4. . Directive 2011/65/EU of the European Parliment and of the Council of 8 June 2011 on the Restriction of the use of Certain Hazardous Substances in Electrical and Electronic Equipment 2024.
5. Jang, E.; Kim, Y.; Won, Y.-H.; Jang, H.; Choi, S.-M., Environmentally Friendly InP-Based Quantum Dots for Efficient Wide Color Gamut Displays. *ACS Energy Lett.* **2020**, *5* (4), 1316-1327.
6. Chopra, S. S.; Theis, T. L., Comparative Cradle-to-Gate Energy Assessment of Indium Phosphide and Cadmium Selenide Quantum Dot Displays. *Environ. Sci. Nano* **2017**, *4* (1), 244-254.
7. Saha, A.; Figueroba, A.; Konstantatos, G., $Ag_2ZnSnS_4$ Nanocrystals Expand the Availability of RoHS Compliant Colloidal Quantum Dots. *Chem. Mater.* **2020**, *32* (5), 2148-2155.
8. Wang, Y.; Kavanagh, S. R.; Burgués-Ceballos, I.; Walsh, A.; Scanlon, D. O.; Konstantatos, G., Cation Disorder Engineering Yields $AgBiS_2$ Nanocrystals with Enhanced Optical Absorption for Efficient Ultrathin Solar Cells. *Nat. Photonics* **2022**, *16* (3), 235-241.
9. Quarta, D.; Toso, S.; Fieramosca, A.; Dominici, L.; Caliandro, R.; Moliterni, A.; Tobaldi, D. M.; Saleh, G.; Gushchina, I.; Brescia, R.; Prato, M.; Infante, I.; Cola, A.; Giannini, C.; Manna, L.; Gigli, G.; Giansante, C., Direct Band Gap Chalcohalide Semiconductors: Quaternary $AgBiSCl_2$ Nanocrystals. *Chem. Mater.* **2023**, *35* (23), 9900-9906.
10. Jang, H.; Jung, Y. S.; Oh, M.-W., Advances in Thermoelectric $AgBiSe_2$: Properties, Strategies, and Future Challenges. *Heliyon* **2023**, *9* (11), e21117.
11. Kwon, J.; Shin, Y.; Sung, Y.; Doh, H.; Kim, S., Silver Sulfide Nanocrystals and Their Photodetector Applications. *Acc. Mater.* **2024**, *5* (9), 1097-1108.
12. Kuhs, W. F.; Nitsche, R.; Scheunemann, K., The Argyrodites — A New Family of Tetrahedrally Close-Packed Structures. *Mater. Res. Bull.* **1979**, *14* (2), 241-248.
13. Lin, S.; Li, W.; Pei, Y., Thermally Insulative Thermoelectric Argyrodites. *Mater. Today* **2021**, *48*, 198-213.
14. Bindi, L.; Nestola, F.; Guastoni, A.; Zorzi, F.; Peruzzo, L.; Raber, T., Te-Rich Canfieldite, $Ag_8Sn(S, Te)_6$, from the Lengenbach Quarry, Binntal, Canton Valais, Switzerland: Occurrence, Description and Crystal Structure. *Canad Mineral* **2012**, *50* (1), 111-118.
15. Heep, B. K.; Weldert, K. S.; Krysiak, Y.; Day, T. W.; Zeier, W. G.; Kolb, U.; Snyder, G. J.; Tremel, W., High Electron Mobility and Disorder Induced by Silver Ion Migration Lead to Good Thermoelectric Performance in the Argyrodite $Ag_8SiSe_6$. *Chem. Mater.* **2017**, *29* (11), 4833-4839.
16. Schnepf, R. R.; Cordell, J. J.; Tellekamp, M. B.; Melamed, C. L.; Greenaway, A. L.; Mis, A.; Brennecka, G. L.; Christensen, S.; Tucker, G. J.; Toberer, E. S.; Lany, S.; Tamboli, A. C., Utilizing Site Disorder in the Development of New Energy-Relevant Semiconductors. *ACS Energy Lett.* **2020**, *5* (6), 2027-2041.
17. Feng, X.; Chien, P.-H.; Wang, Y.; Patel, S.; Wang, P.; Liu, H.; Immediato-Scuotto, M.; Hu, Y.-Y., Enhanced Ion Conduction by Enforcing Structural Disorder in Li-Deficient Argyrodites $Li_{6-x}PS_{5-x}Cl_{1+x}$. *Energy Storage Mater.* **2020**, *30*, 67-73.
18. Kang, S.; Lee, S.; Lee, H.; Kang, Y.-M., Manipulating Disorder Within Cathodes of Alkali-Ion Batteries. *Nat. Rev. Chem.* **2024**, *8* (8), 587-604.
19. Roychowdhury, S.; Ghosh, T.; Arora, R.; Samanta, M.; Xie, L.; Singh, N. K.; Soni, A.; He, J.; Waghmare, U. V.; Biswas, K., Enhanced Atomic Ordering Leads to High Thermoelectric Performance in $AgSbTe_2$. *Science* **2021**, *371* (6530), 722-727.
20. Palache, C., Memorial of Frederick Alexander Canfield. *Am. Mineral.* **1927**, *12* (3), 67-70.
21. Mikolaichuk, A. G.; Moroz, N. V.; Demchenko, P. Y.; Akselrud, L. G.; Gladyshevskii, R. E., Phase Relations in the $Ag_8SnS_6$-$Ag_2SnS_3$-$AgBr$ System and Crystal Structure of $Ag_6SnS_4Br_2$. *Inorg. Mater.* **2010**, *46* (6), 590-597.
22. Shen, X.; Xia, Y.; Yang, C.-C.; Zhang, Z.; Li, S.; Tung, Y.-H.; Benton, A.; Zhang, X.; Lu, X.; Wang, G.; He, J.; Zhou, X., High Thermoelectric Performance in Sulfide-Type Argyrodites Compound $Ag_8Sn(S_{1-x}Se)_6$ Enabled by Ultralow Lattice Thermal Conductivity and Extended Cubic Phase Regime. *Adv. Funct. Mater.* **2020**, *30* (21), 2000526.
23. Kameyama, T.; Fujita, S.; Furusawa, H.; Torimoto, T., Size-Controlled Synthesis of $Ag_8SnS_6$ Nanocrystals for Efficient Photoenergy Conversion Systems Driven by Visible and Near-IR Lights. *Part. Syst. Charact.* **2014**, *31* (11), 1122-1126.
24. Zhu, L., Xu, Yafeng, Zheng, Haiying, Liu, Guozhen, Xu, Xiaoxiao, Pan, Xu, Dai, Songyuan, Application of Facile Solution-Processed Ternary Sulfide $Ag_8SnS_6$ as Light Absorber in Thin Film Solar Cells. *Sci. China Mater.* **2018**, *61* (12), 1549-1556.
25. Takahashi, S.; Kasai, H.; Liu, C.; Miao, L.; Nishibori, E., Rattling of Ag Atoms Found in the Low-Temperature Phase of Thermoelectric Argyrodite $Ag_8SnSe_6$. *Cryst. Growth Des.* **2024**, *24* (15), 6267-6274.
26. Li, Z.; Li, W.; Shao, H.; Dou, M.; Cheng, Y.; Wan, X.; Jiang, X.; Zhang, Z.; Chen, Y.; Li, S., Water-Soluble Ag–Sn–S Nanocrystals Partially Coated with ZnS Shells for Photocatalytic Degradation of Organic Dyes. *ACS Appl. Nano Mater.* **2023**, *6* (6), 4417-4427.
27. Slade, T. J.; Gvozdetskyi, V.; Wilde, J. M.; Kreyssig, A.; Gati, E.; Wang, L.-L.; Mudryk, Y.; Ribeiro, R. A.; Pecharsky, V. K.; Zaikina, J. V.; Bud'ko, S. L.; Canfield, P. C., A Low-Temperature Structural Transition in Canfieldite, $Ag_8SnS_6$, Single Crystals. *Inorg. Chem.* **2021**, *60* (24), 19345-19355.
28. Katty, A.; Gorochov, O.; Letoffe, J. M., Etude Radiocristallographique et Calorimetrique des Transitions de Phase de $Ag_8GeTe_6$. *J. Solid State Chem.* **1981**, *38* (2), 259-263.
29. Tan, J. M. R.; Lee, Y. H.; Pedireddy, S.; Baikie, T.; Ling, X. Y.; Wong, L. H., Understanding the Synthetic Pathway of a Single-Phase Quarternary Semiconductor Using Surface-Enhanced Raman Scattering: A Case of Wurtzite $Cu_2ZnSnS_4$ Nanoparticles. *J. Am. Chem. Soc.* **2014**, *136* (18), 6684-6692.
30. Chang, J.; Waclawik, E. R., Controlled Synthesis of $CuInS_2$, $Cu_2SnS_3$ and $Cu_2ZnSnS_4$ Nano-Structures: Insight into the Universal Phase-Selectivity Mechanism. *CrystEngComm* **2013**, *15* (28), 5612-5619.
31. Yarur Villanueva, F.; Green, P. B.; Qiu, C.; Ullah, S. R.; Buenviaje, K.; Howe, J. Y.; Majewski, M. B.; Wilson, M. W. B., Binary $Cu_{2-x}S$ Templates Direct the Formation of Quaternary $Cu_2ZnSnS_4$ (Kesterite, Wurtzite) Nanocrystals. *ACS Nano* **2021**, *15* (11), 18085-18099.
32. Yu, K.; Carter, E. A., A Strategy to Stabilize Kesterite CZTS for High-Performance Solar Cells. *Chem. Mater.* **2015**, *27* (8), 2920-2927.
33. Guin, S. N.; Biswas, K., Cation Disorder and Bond Anharmonicity Optimize the Thermoelectric Properties in Kinetically Stabilized Rocksalt $AgBiS_2$ Nanocrystals. *Chem. Mater.* **2013**, *25* (15), 3225-3231.
34. Öberg, V. A.; Johansson, M. B.; Zhang, X.; Johansson, E. M. J., Cubic $AgBiS_2$ Colloidal Nanocrystals for Solar Cells. *ACS Appl. Nano Mater.* **2020**, *3* (5), 4014-4024.
35. Yarur Villanueva, F.; Hasham, M.; Green, P. B.; Imperiale, C. J.; Rahman, S.; Burns, D. C.; Wilson, M. W. B., A Stepwise Reaction Achieves Ultrasmall $Ag_2ZnSnS_4$ Nanocrystals. *ACS Nano* **2024**, *18* (52), 35182-35201.
36. Ning, J.; Zou, B., Controlled Synthesis, Formation Mechanism, and Applications of Colloidal $Ag_8SnS_6$ Nanoparticles and $Ag_8SnS_6$/$Ag_2S$ Heterostructured Nanocrystals. *J. Phys. Chem. C* **2018**, *122* (12), 6566-6572.
37. Sasamura, T.; Osaki, T.; Kameyama, T.; Shibayama, T.; Kudo, A.; Kuwabata, S.; Torimoto, T., Solution-Phase Synthesis of Stannite-Type $Ag_2ZnSnS_4$ Nanoparticles for Application to Photoelectrode Materials. *Chem. Lett.* **2012**, *41* (9), 1009-1011.
38. Park, N.; Friedfeld, M. R.; Cossairt, B. M., Chapter 5 - Semiconductor Clusters and their use as Precursors to Nanomaterials. In *Nanomaterials via Single-Source Precursors*, Apblett, A. W.; Barron, A. R.; Hepp, A. F., Eds. Elsevier: 2022; pp 165-200.

39. Friedfeld, M. R.; Stein, J. L.; Ritchhart, A.; Cossairt, B. M., Conversion Reactions of Atomically Precise Semiconductor Clusters. *Acc. Chem. Res.* **2018,** *51* (11), 2803-2810.

40. Busatto, S.; de Mello Donega, C., Magic-Size Semiconductor Nanostructures: Where Does the Magic Come from? *ACS Mater Au* **2022,** *2* (3), 237-249.

41. Nagaoka, Y.; Tan, R.; Li, R.; Zhu, H.; Eggert, D.; Wu, Y. A.; Liu, Y.; Wang, Z.; Chen, O., Superstructures Generated from Truncated Tetrahedral Quantum Dots. *Nature* **2018,** *561* (7723), 378-382.

42. Shanbhag, S.; Kotov, N. A., On the Origin of a Permanent Dipole Moment in Nanocrystals with a Cubic Crystal Lattice: Effects of Truncation, Stabilizers, and Medium for CdS Tetrahedral Homologues. *J. Phys. Chem. B* **2006,** *110* (25), 12211-12217.

43. Leemans, J.; Dümbgen, K. C.; Minjauw, M. M.; Zhao, Q.; Vantomme, A.; Infante, I.; Detavernier, C.; Hens, Z., Acid–Base Mediated Ligand Exchange on Near-Infrared Absorbing, Indium-Based III–V Colloidal Quantum Dots. *J. Am. Chem. Soc.* **2021,** *143* (11), 4290-4301.

44. Patterson, A. L., The Scherrer Formula for X-Ray Particle Size Determination. *Physical Review* **1939,** *56* (10), 978-982.

45. Young, N. P.; van Huis, M. A.; Zandbergen, H. W.; Xu, H.; Kirkland, A. I., Transformations of Gold Nanoparticles Investigated Using Variable Temperature High-Resolution Transmission Electron Microscopy. *Ultramicroscopy* **2010,** *110* (5), 506-516.

46. Thomas, E. M.; Pradhan, N.; Thomas, K. G., Reasoning the Photoluminescence Blinking in CdSe–CdS Heteronanostructures as Stacking Fault-Based Trap States. *ACS Energy Lett.* **2022,** *7* (8), 2856-2863.

47. Majumder, S.; Bae, I.-T.; Maye, M. M., Investigating the Role of Polytypism in the Growth of Multi-Shell CdSe/CdZnS Quantum Dots. *J. Mater. Chem. C* **2014,** *2* (23), 4659-4666.

48. Holder, C. F.; Schaak, R. E., Tutorial on Powder X-Ray Diffraction for Characterizing Nanoscale Materials. ACS Nano: 2019; Vol. 13, pp 7359-7365.

49. Cowley, J. M., Diffraction Physics. 3rd rev. ed ed.; Elsevier Science B.V.: Amsterdam, 1995. http://site.ebrary.com/id/10190073.

50. Peng, S.; Lee, Y.; Wang, C.; Yin, H.; Dai, S.; Sun, S., A Facile Synthesis of Monodisperse Au Nanoparticles and their Catalysis of CO Oxidation. *Nano Research* **2008,** *1* (3), 229-234.

51. Wang, N., New Data for $Ag_8SnS_6$ (Canfieldite) and $Ag_8GeS_6$ (Argyrodite). *Neues Jahrb. Mineral., Monatsh.* **1978,** 269–272.

52. Scholz, T.; Schneider, C.; Terban, M. W.; Deng, Z.; Eger, R.; Etter, M.; Dinnebier, R. E.; Canepa, P.; Lotsch, B. V., Superionic Conduction in the Plastic Crystal Polymorph of $Na_4P_2S_6$. *ACS Energy Lett.* **2022,** *7* (4), 1403-1411.

53. Dzhagan, V.; Selyshchev, O.; Havryliuk, Y.; Mazur, N.; Raievska, O.; Stroyuk, O.; Kondratenko, S.; Litvinchuk, A. P.; Valakh, M. Y.; Zahn, D. R. T., Raman and X-ray Photoelectron Spectroscopic Study of Aqueous Thiol-Capped Ag-Zn-Sn-S Nanocrystals. *Materials* **2021,** *14* (13), 3593.

54. Pietak, K.; Jastrzebski, C.; Zberecki, K.; Jastrzebski, D. J.; Paszkowicz, W.; Podsiadlo, S., Synthesis and Structural Characterization of $Ag_2ZnSnS_4$ crystals. *J. Solid State Chem.* **2020,** *290*, 121467.

55. Zhu, D.; Bellato, F.; Bahmani Jalali, H.; Di Stasio, F.; Prato, M.; Ivanov, Y. P.; Divitini, G.; Infante, I.; De Trizio, L.; Manna, L., $ZnCl_2$ Mediated Synthesis of InAs Nanocrystals with Aminoarsine. *J. Am. Chem. Soc.* **2022,** *144* (23), 10515-10523.

56. McVey, B. F. P.; Swain, R. A.; Lagarde, D.; Tison, Y.; Martinez, H.; Chaudret, B.; Nayral, C.; Delpech, F., Unraveling the Role of Zinc Complexes on Indium Phosphide Nanocrystal Chemistry. *J. Chem. Phys.* **2019,** *151* (19), 191102.

57. Villanueva, F. Y. Understanding and Controlling the Formation Mechanism and Surface Chemistry of Lead-Free Quaternary Semiconductor Nanocrystals. University of Toronto (Canada), ProQuest, 2024.

58. Saniepay, M.; Mi, C.; Liu, Z.; Abel, E. P.; Beaulac, R., Insights into the Structural Complexity of Colloidal CdSe Nanocrystal Surfaces: Correlating the Efficiency of Nonradiative Excited-State Processes to Specific Defects. *J. Am. Chem. Soc.* **2018,** *140* (5), 1725-1736.

59. Houtepen, A. J.; Hens, Z.; Owen, J. S.; Infante, I., On the Origin of Surface Traps in Colloidal II–VI Semiconductor Nanocrystals. *Chem. Mater.* **2017,** *29* (2), 752-761.

TOC

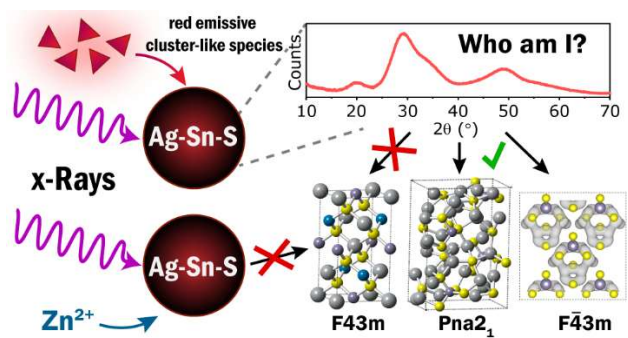

# Supplementary Information

# Synthesis and Structural Analysis of an Emissive Colloidal Argyrodite Nanocrystal: Canfieldite $Ag_8SnS_6$

Francisco Yarur Villanueva,[1,2] Victor Quezada Novoa,[3] Pascal Rusch,[1] Stefano Toso,[1] Maxwell W. Terban,[4] Yurii P. Ivanov,[1] Joaquin Carlos Chu,[2] Maxine J. Kirshenbaum,[2] Ehsan Nikbin,[5] [1] Maria J. Gendron Romero,[2] Mirko Prato,[1] Giorgio Divitini,[1] Jane Y. Howe,[5] Mark W.B. Wilson,[2*] Liberato Manna[1*]

[1]Istituto Italiano di Tecnologia, Via Morego 30, 16163, Genova, Italy

[2]Department of Chemistry, University of Toronto, Toronto, Ontario, M5S 3H6, Canada

[3]Department of Chemistry and Biochemistry and Centre for NanoScience Research, Concordia University, 7141 Sherbrooke Street West, Montréal, Quebec H4B 1R6, Canada

[4]Momentum Transfer GmbH, Luruper Hauptstraße 1, 22547 Hamburg, Germany

[5]University of Toronto, Department of Materials Science and Engineering, Toronto, Ontario, M5S3E4, Canada

Contents

### Section 1. Methods

*Chemicals:* Silver (I) Chloride [AgCl], Tin (IV) Chloride [$SnCl_4$], Zinc (II) Bromide [$ZnBr_2$], hexamethyldisilathiane [$(TMS)_2$-S], Oleylamine [OLA 98%], Toluene 99%, Ethanol [EtOH, 99.9%], Methanol (99%), Hexane [≥99%], dimethylformamide [DMF  98%], and tetrabutylammonium bromide (TBABr, 98%) were purchased from Sigma Aldrich. All chemicals were used as received without further purification except for OLA, which was degassed at 125 °C under vacuum for 3 hours and stored under argon.

*Synthesis of ATS and ATS@Zn-1 Nanocrystals (adapted from Yarur Villanueva et al.)[1]:* A 25 mL 3-neck round-bottom flask was loaded with AgCl (42 mg, 0.30 mmol), sealed with a rubber septum and connected to a Schlenk line *via* condenser. The flask was evacuated, and 8 mL of degassed OLA 98% were added while stirring. This flask is placed under vacuum and heated to 70 °C for 1 hr for the AgCl to dissolve.

While the AgCl precursor dissolved, one 20 mL scintillation vial was loaded with degassed OLA 98% (1.5 mL) and a stir bar and brought inside a $N_2$ glovebox to prepare the $SnCl_4$ precursor solution. A micropipette was used to add $SnCl_4$ (18 μL, 0.15 mmol) to this vial. It is important to tilt the vial (~45°) and deliver the $SnCl_4$ onto a dry spot at the bottom of the vial to prevent traces of water in the OLA from hydrolyzing inside of the pipette tip and forming an inconvenient gel. For the synthesis of ATS@Zn NCs, another 20 mL scintillation vial was charged with $ZnBr_2$ (34 mg, 0.15 mmol) and degassed OLA 98% (1.5 mL) inside a $N_2$ glovebox. Both vials containing the $SnCl_4$ and $ZnBr_2$ precursors were taken outside of the glove box and then placed in a sand bath at ~70 °C for 30 min. When all metal precursors dissolved, two $(TMS)_2$-S solutions were prepared in different 20 mL scintillation vials inside the glove box. The first solution [1] contained toluene (1.5 mL) and $(TMS)_2$-S (26 μL/0.122 mmol), while the second [2] was comprised of toluene (1.5 mL) and $(TMS)_2$-S (98 μL/0.464 mmol). Each solution was loaded into a labelled 3 mL plastic syringe and taken out of the glovebox. This completed the precursor preparation.

The reaction temperature is set to 70 °C and the injection set up is assembled. A 3 mL plastic syringe is loaded with the tin solution (Sn-OLA) (a second syringe is loaded with the zinc precursor (Zn-OLA) for ATS@Zn NCs), both fully dissolved and pale yellow in colour. An additional 14/20 rubber septum is pierced with all four syringes as seen in Video 1. Then, the existing septum on the flask is removed under a positive nitrogen flow. This septum is quickly replaced by the one holding the syringes with the precursor solutions. The injection occurs at 70 °C in the order: 1) $(TMS)_2$-S [1], 2) Sn-OLA, and 3) $(TMS)_2$-S [2] with one second between injections. (a fourth injection 4) Zn-OLA is made for ATS@Zn NCs). The flask is removed from the heating mantle and rapidly cooled by blowing cold air. See attached videos for a full demonstration of the injection procedure.

To achieve larger sizes (>5 nm), the precursor injection was done at 80 °C and the reaction temperature was set to 95 °C. Aliquots for TEM analysis were taken after 10 minutes at 95 °C.

NCs are purified through a standard EtOH workup. Essentially, the crude solution (14 mL) is transferred and split into two 50 mL Falcon tubes and 14 mL of EtOH are added to each tube. The tubes are centrifuged at 4430 rfc for 3 minutes. The pale-yellow supernatant is discarded and the NCs are re-dispersed in 1 mL hexanes. Then, ~700 μL of EtOH are added and the solution is centrifuged at 4430 rfc for 30 seconds. The NC pellet is dispersed in 1 mL hexanes for further experiments.

*Size-selective precipitation*: Using a micropipettor, 5 mL of crude solution were placed in a 50 mL Falcon tube. 7.5 mL of EtOH (99%) were added and the solution was centrifuged at 4430 rfc for 3 minutes. The pellet was re-dispersed in 500 μL of hexanes and the supernatant was transferred into another Falcon tube to which 1.5 mL of EtOH were added. The solution was centrifuged at 4430 rfc for 2 minutes. Then, the pellet was re-dispersed in 500 μL hexanes and the supernatant was collected into another Falcon tube and 1.5 mL EtOH were added, followed by centrifugation at 4430 rfc for 2 minutes. The pellet was re-dispersed in 500 μL hexanes and the supernatant was transferred into another Falcon tube. EtOH was added until the 22.5 mL mark on the tube, followed by centrifugation at 4430 rfc for 2 minutes.

To isolate and stabilize the ATS cluster-like species, we added 1.5 mL of a 0.2 M $ZnBr_2$ in EtOH and 6 mL of pure EtOH instead of 7.5 mL of EtOH in the first purification step. All other steps were performed as described above. Cluster-like species are recovered in the third or fourth supernatant.

*Synthesis of reported ATS@Zn-2 NCs*: These NCs were synthesized following the procedure reported by Saha *et al*. at 160 °C for 2 hrs.[7]

*Purification procedure for HR-TEM*: NC samples were purified through a regular two-step purification with ethanol and hexanes (See above). Then, 100 µL of the 1 mL stock solution were diluted in 200 µL hexanes and this dispersion was placed in a vial containing 500 µL of a 0.075 M TBABr solution in DMF to create a bi-phasic system. 5 µL of butylamine were added into the hexanes phase and the vial was closed and shaken for 10 seconds for the NCs to transfer from the hexanes to the DMF phase. The hexanes phase was removed with a pipette and 8-10 mL of toluene were added to precipitate the NCs *via* centrifugation for 30 seconds at 4430 rfc. The pellet was re-dispersed in 75 µL of DMF and toluene was carefully added dropwise until NC precipitation was observed (adding too much toluene will cause aggressive aggregation and the pellet will not be able to re-disperse for TEM analysis). The solution was centrifuged for 10 seconds at 4430 rfc, the supernatant was discarded, and the NCs were re-dispersed in 100 µL of DMF to create a stock solution. ~20 µL of this stock solution were diluted in 400 µL DMF to create a solution for drop casting onto TEM grids.

*Characterization:*

*Transmission Electron Microscopy (TEM):* BF TEM images were acquired using a Hitachi HT7700 microscope at 100 kV. 400-mesh copper grids (Pacific grid) were immersed in a dilute dispersion of NCs in hexane for 1 second and left to dry under air. The average size was determined over an average of 250-400 particles using the Fiji imaging-processing distribution of the ImageJ2 software. High-resolution measurements were acquired using either a Hitachi HF3300 equipped with a cold field emission electron gun, operated at 300 kV or a probe- corrected Thermo Fisher Spectra 300 S/TEM operated at 300 kV, employing a HAADF detector with a beam current of a few tens of picoamperes to limit beam damage to the sample. Compositional maps were acquired using rapid raster scanning in Velox, with a probe current of ~150 pA. Elemental maps were produced after re-binning and local averaging within Velox.

*Steady-State Photoluminescence (PL):* Photoluminescence spectra taken with a home-built set-up at λ=450 nm excitation using a pen-diode (ThorLabs CPS450) at 2.5 mW. The emission set-up involved an off-axis parabolic collimating mirror to direct the emission from the sample to a reflective fibre-coupler (ThorLabs PC12FC-P01), which was then sent to an OceanOptics Flame spectrometer. The PLQY was measured by using a calibrated integrating sphere in a FS5 Spectrofluorimeter, Edinburgh Instruments ($\lambda_{ex}$=500 nm).

*Absorption spectroscopy (UV-Vis):* Optical absorption spectra were taken on an Agilent Cary 5000 UV/Vis spectrophotometer.

*X-Ray Diffraction:* XRD analysis was performed on a PANanalytical Empyrean X-ray diffractometer, equipped with a 1.8 kW Cu Kα ceramic anode and a PIXcel3D 2 × 2 area detector, operating at 45 kV and 40 mA. The samples were drop cast from hexanes onto a silicon substrate and scanned for 6 to 24 hrs.

*Scanning Electron Microscopy Energy Dispersive Spectroscopy (SEM-EDS)*: Three drops of a 120 mg/mL stock NC dispersion in hexanes were drop cast onto an amorphous silicon substrate and dried under ambient conditions overnight. SEM-EDS measurements were performed using a JEOL Dry SD30GV silicon-drift detector (SDD), with 30 mm$^2$ effective area.

*X-ray photoelectron spectroscopy (XPS):* A 40 mg/mL ATS dispersion in hexanes was purified through five standard cycles of precipitation using EtOH and centrifugation, with re-dispersion in hexanes at each step. For each of these washes, the pellet was re-dispersed in 1 mL hexanes and 1 mL EtOH was added to induce precipitation of the NCs, followed by centrifugation at 3904 rfc for 1 min in each cycle. The clear supernatant was discarded after each centrifugation and the final pellet was dried under vacuum overnight.

Measurements were carried out using a Kratos Axis UltraDLD spectrometer (Kratos Analytical Ltd.) with a monochromated Al Kα X-ray source (hν = 1486.6 eV) operating at 20 mA and 15 kV. Each sample was grounded to the sample holder *via* copper tape to maximize its electrical conductivity. Wide-area scans were collected over an analysis area of 300 × 700 µm$^2$ at a photoelectron pass energy of 160 eV and energy step of 1 eV, while high-resolution spectra were collected at a photoelectron pass energy of 20 eV and an energy step of 0.1 eV. A take-off angle of 0° with respect to sample normal direction was used for all analyses. The differential electrical charging effects were neutralized. The spectra have been referenced to the adventitious carbon 1s peak at 284.8 eV. The spectra were analyzed with the CasaXPS software (Casa Software Ltd., version 2.3.24)[2] and the residual background was eliminated by the Shirley method.

*High-resolution powder X-ray diffraction and Pair distribution function analysis (PDF)*: High-resolution synchrotron X-ray diffraction and total scattering measurements were performed at beamline ID31 at the European Synchrotron Radiation Facility (ESRF). The sample powders were loaded into cylindrical slots (approx. 1 mm thickness) held between Kapton windows in a high-throughput sample holder. Each sample was measured in a transmission with an incident X-ray energy of 75.00 keV (λ = 0.1653 Å). Measured intensities were collected using a Pilatus CdTe 2M detector (1679 × 1475 pixels, 172 × 172 µm$^2$ each) positioned with the incident beam in the corner of the detector. The sample-to-detector distance was approximately 1.5 m for the high-resolution measurements and 0.3 m for the total scattering measurement. Background measurements for the empty windows were measured and subtracted. NIST SRM 660b (LaB6) was used for geometry calibration performed with the software pyFAI followed by image integration including a flat-field, geometry, solid-angle, and polarization corrections.

High-resolution PXRD data were background subtracted. Preliminary PDF data were processed in an automated way using PDFgetX3 with $Q_{max}$ = 25 Å$^{-1}$. Then the data were reprocessed using a Lorch modification function to suppress termination effects and contributions from high frequency noise. The small-angle scattering intensities were extrapolated from $Q_{min}$ = 0.328 Å$^{-1}$ to 0. The compositions of the sample were taken into consideration to process de PDFs.

**Section 2. Physical and Optical Characterization for ATS NCs**

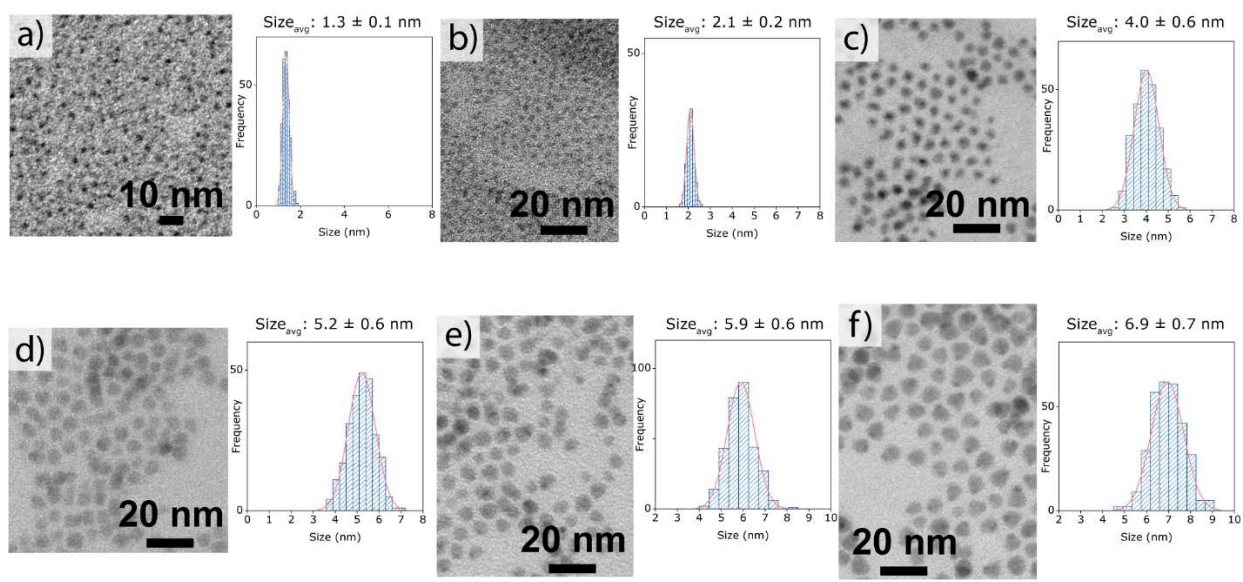

**Figure S 1**Error! Bookmark not defined.. BF TEM images and histograms showing size control for ATS (b to f) and ATS cluster-like species (a) NCs. The size of these NCs can be tuned in the 1.3 to 6.9 nm range.

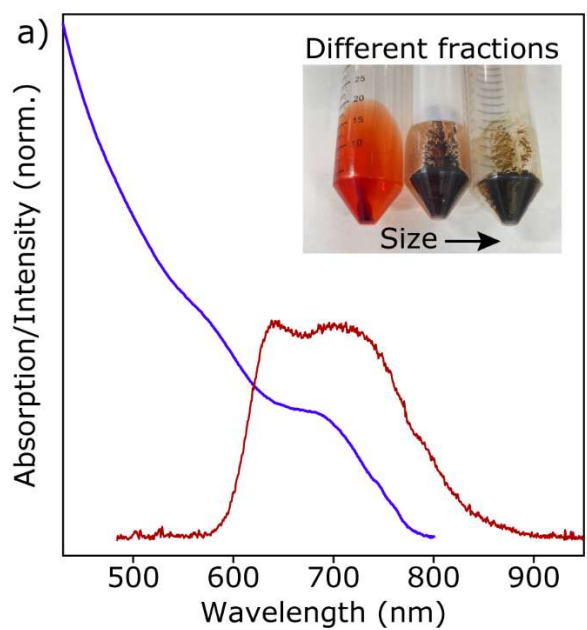

**Figure S 2**. Optical data UV-Vis (blue trace) and photoluminescence (red trace) for the whole product of a typical ATS NCs reaction. Various features can be observed in either spectra due to polydispersity.

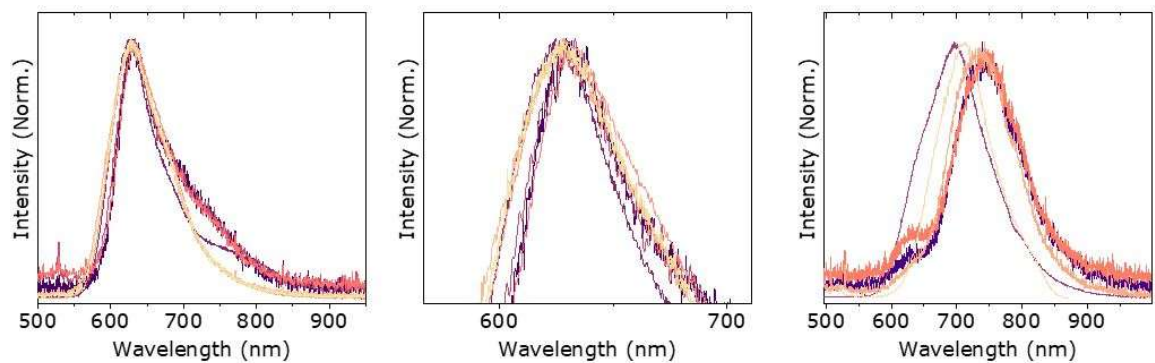

**Figure S 3.** a) PL spectra showing the reproducibility and consistency in emission (peak at ~630 nm) for ATS cluster-like species over 7 syntheses performed in different days and laboratories (The University of Toronto and the Istituto Italiano di Tecnologia). b) Peak zoomed in of (a), and c) PL spectra for larger ATS NCs shows emission variability.

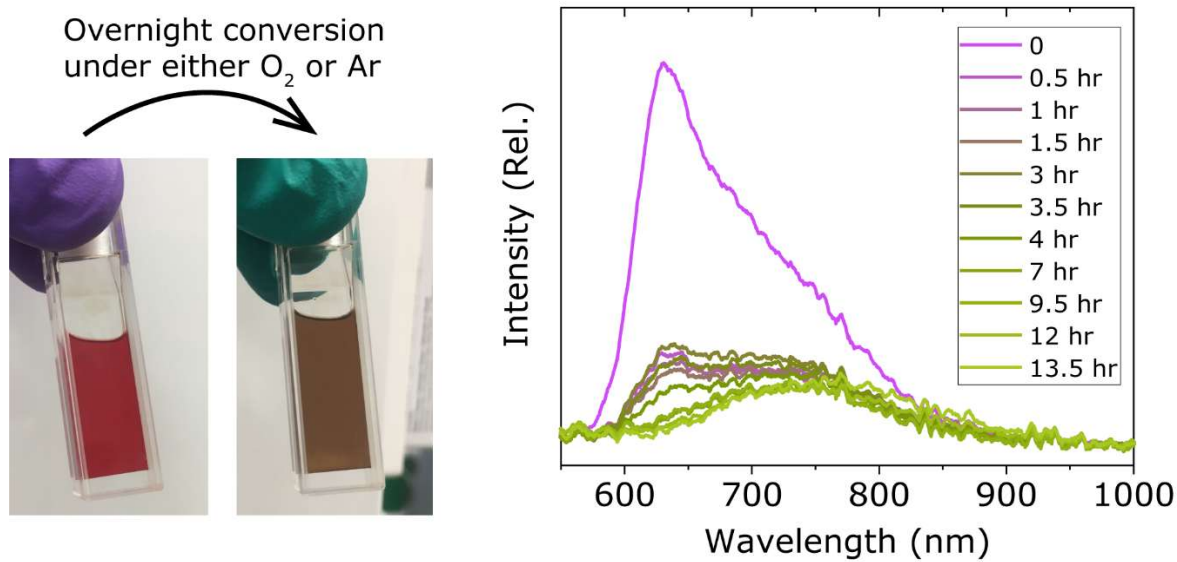

**Figure S 4.** Cluster-like ATS species with emission peak at ~630 nm converts to larger ATS NCs overnight under constant 450 nm laser irradiation (~2 mW at the focal point). While a continue laser was used here to enable the acquisition of the time-series of emission spectra, we do not consider that laser excitation is necessary for the conversion. This experiment was performed under an argon atmosphere by purging the solution in a sealed cuvette for 2 minutes and leaving it at a positive argon pressure *via* a ballon.

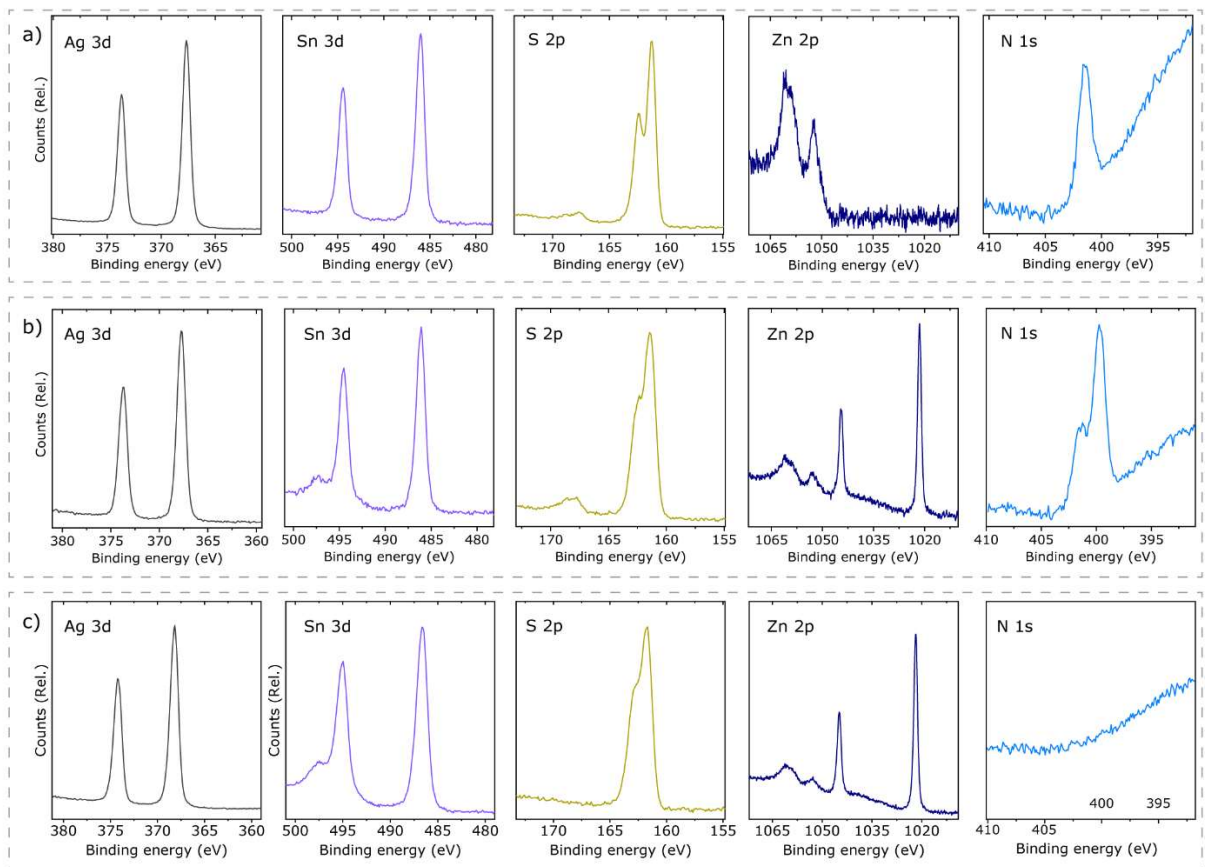

**Figure S 5**. X-ray photoelectron spectroscopy data for a) ATS, b) ATS@Zn-1, and c) ATS@Zn-2 NC samples of average size 6.0 nm. The most relevant observations in this set of data concerns elements Sn, S, and N (See XPS discussion below).

Sn: Quantitative analysis suggests that tin is present as Sn (IV) in both systems. However, determination of the oxidation state of Sn is not straightforward, as Sn (II) and Sn (IV) signals often overlap. Typically, the analysis requires the combination of XPS and Auger data to obtain a Wagner plot.[3] Such analysis on ATS and ATS@Zn samples suggests that Sn is in a 2+, contrary to our quantitative analysis *via* XPS, which might indicate that some Sn (II) is present at the surface where it is more prone to redox reactions. Although the behaviour of Sn in terms of oxidation states in the lattice is interesting, it is out of the scope of this study.

S: The doublet fitting requires three different components, which might suggest that there is some sulfur oxidation at the surface (*e.g.,* sulfone).

N: A second nitrogen component appears when Zn is present in the reaction. This observation indicates that there are two distinct coordination environments for OLA on the surface. Similarly, it aligns with our previous observations into the surface chemistry of these class of NCs where we observed weaker OLA-metal sites as well as tightly bound $Zn(OLA)_x$ ligands on surface sulfur atoms.[4]

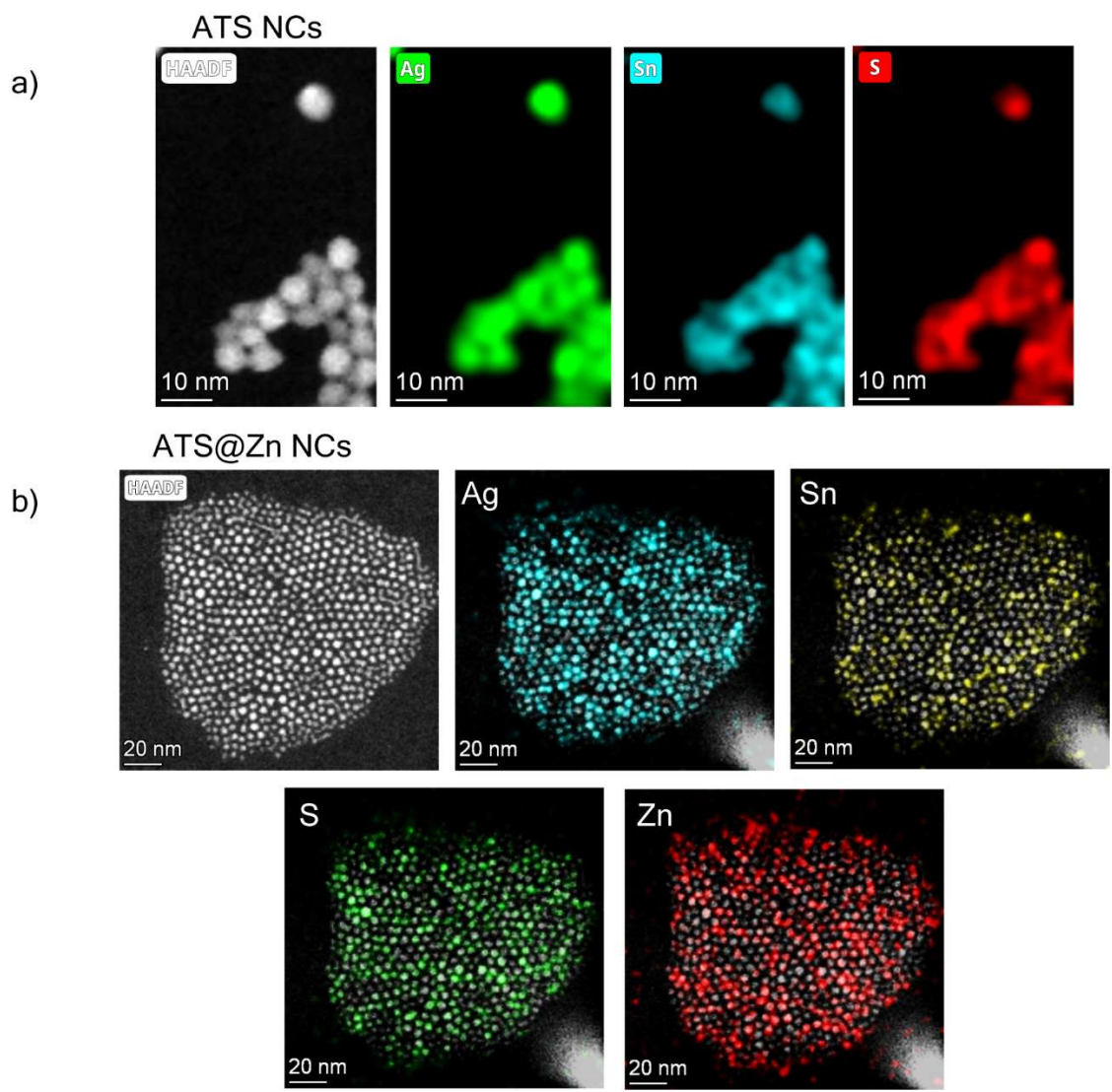

**Figure S 6**. Elemental maps via STEM-EDX of a) ATS and b) ATS@Zn-1 NCs.

| Sample | Technique | Ag | Sn | S | Zn |
|---|---|---|---|---|---|
| **ATS** | XPS | 3.3 | 1.0 | 3.6 | - |
| | STEM-EDX | 4.0 | 1.0 | 3.2 | - |
| **ATS@Zn-1** | XPS | 2.2 | 1.0 | 3.9 | 0.7 |
| | STEM-EDX | 3.9 | 1.0 | 4.8 | 2.1 |
| | SEM-EDS | 2.3 | 1.0 | 4.1 | 1.4 |
| **ATS@Zn-2** | XPS | 2.6 | 1.0 | 3.4 | 0.8 |

**Table S1**. Stoichiometry data acquired through different elemental analysis techniques.

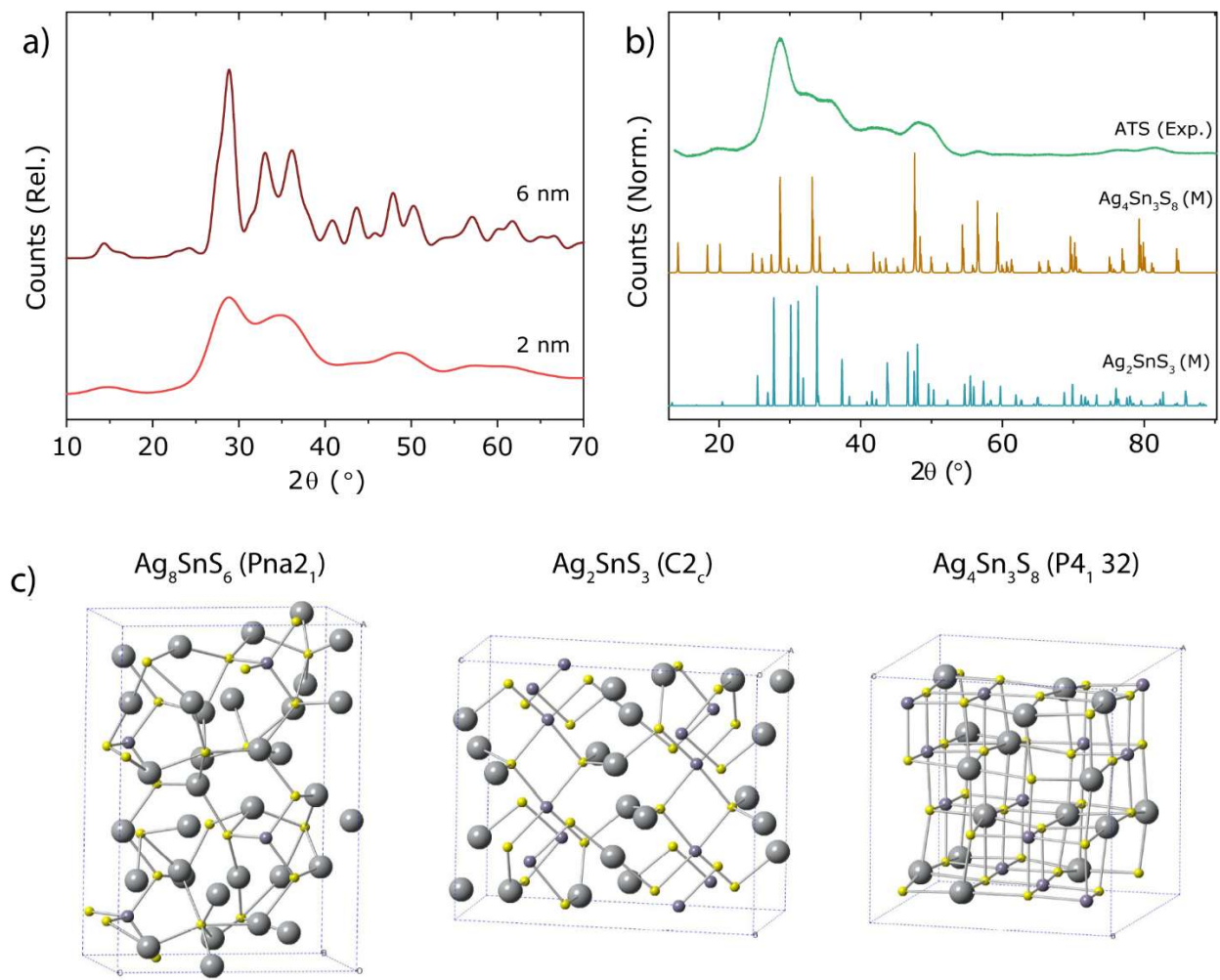

**Figure S 7**. Crystallographic data for ATS and ATS@Zn-1 NCs. a) Simulated PXRD diffractograms, using Crystal Diffract, for Canfieldite with crystal sizes of 2 and 6 nm. b) High-resolution PXRD of ATS NCs (~6.5 nm, green trace) compared to two monoclinic phases in the Ag-Sn-S system, PDF 00-038-0245 ($Ag_4Sn_3S_8$) and PDF 00-039-0140 ($Ag_2SnS_3$). c) Unit cell models for orthorhombic Canfieldite, monoclinic $Ag_2SnS_3$, and monoclinic $Ag_4Sn_3S_8$.

*Rietveld refinement details (Figures 1c and d)*

Simple Rietveld refinements were performed to both the cubic and orthorhombic models to the background subtracted diffraction patterns to see how plausible the indexing of either model is to the observed diffraction features.

In the fits, the lattice parameters were kept constant since the diffraction features are extremely broadened, resulting in a very low sensitivity to lattice parameter values, and to avoid any unrealistic distortions away from the published structure models. Only a scale factor, Gaussian crystallite size broadening term, and the Stephens model using either cubic or orthorhombic setup to allow for anisotropic broadening of the reflections from the crystal models, were refined to fit the peak shapes. The background was described using a Chebychev polynomial of 7th order, and constrained between both refinements so that the resulting fits could be sensibly compared.

**Section 3. Supporting PXRD, TEM, and electron diffraction data**

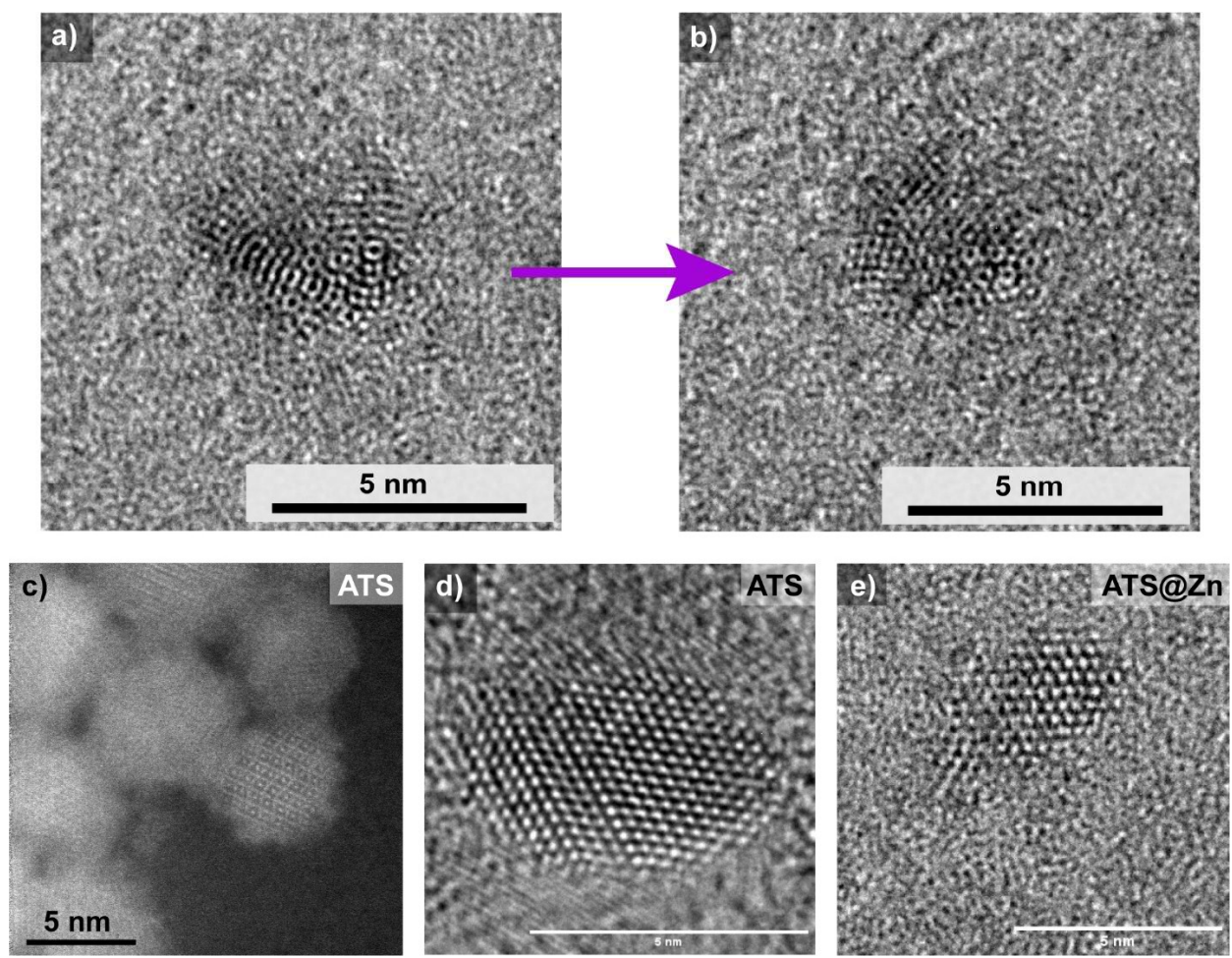

**Figure S 8**. HR-(S)TEM images for ATS and ATS@Zn-1 NCs. a and b) show the coalescence of polycrystalline NCs into single domains under the electron beam. c and d) Single-phase ~5 nm ATS NC displaying stacking faults. e) Single-phased anisotropic ATS@Zn-1 NC.

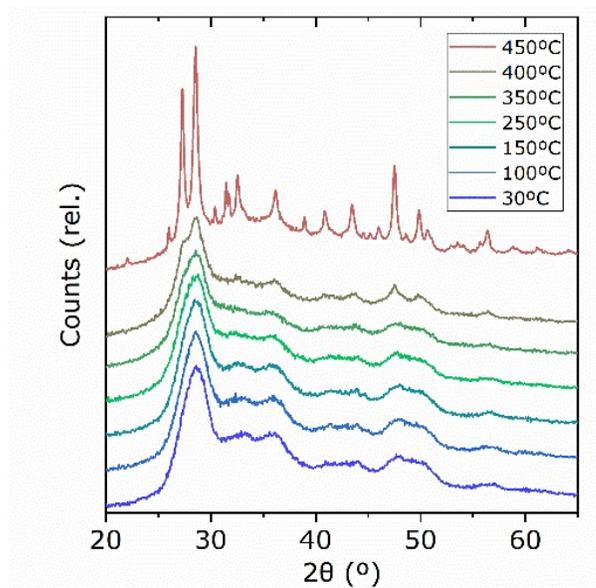

**Figure S 9**. Temperature-dependent PXRD measurements for ATS NCs (~5 nm) on an amorphous silicon substrate. The measurements show little change in the pattern until ~400 °C in which more reflections can be recognize due to crystallite domain growth.

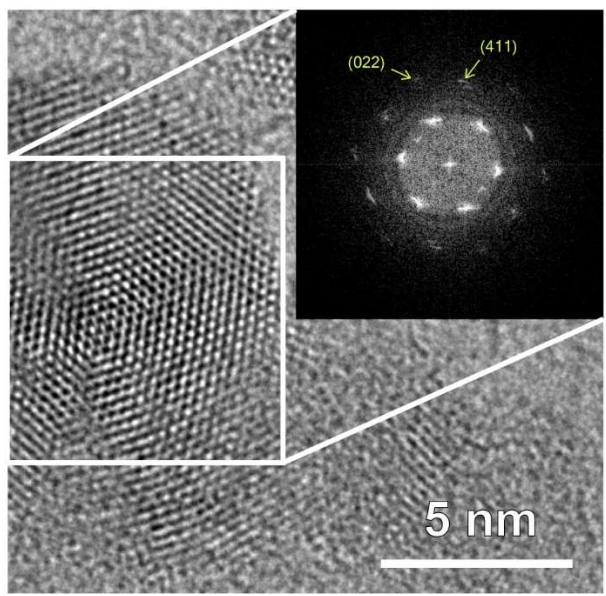

**Figure S 10**. HR-TEM image and corresponding FFT pattern for an ATS NC. The planes (022) and (411) correlate with d-spacing values 0.311 and 0.301 nm matching the orthorhombic phase of canfieldite. (*c.f.* Figure 3)

**Section 4**. PDF analysis.

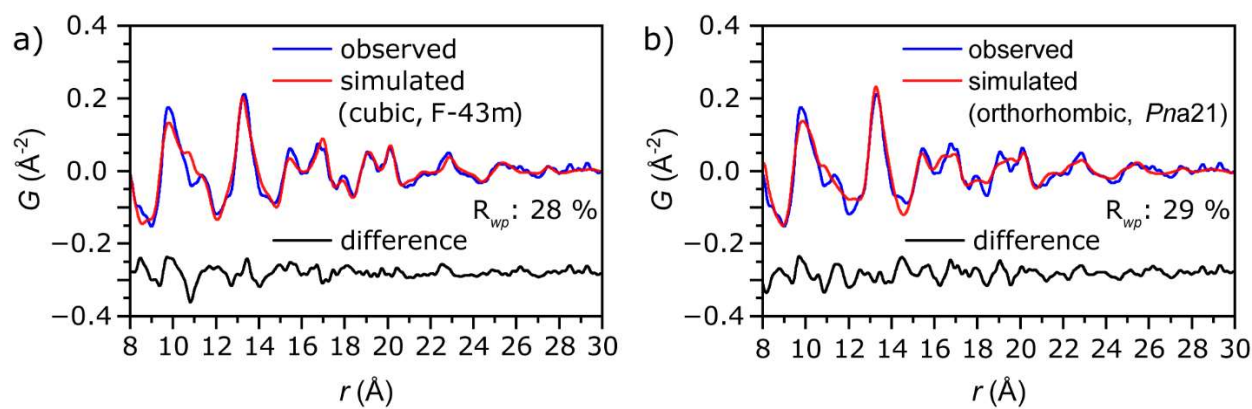

**Figure S 11**. Results of real-space fits of the unmodified cubic (a) and orthorhombic (b) structures to the intermediate range structure observed by the measured PDF (fitted with PDFgui).

*Modelling information*

The models were developed using the TOPAS v7 software using the parameters described in Table S2. Further parameters included lattice parameters, scale factor, isotropic ADPs for Ag, S, and Sn, an intermolecular ADP for SnS$_4$ tetrahedra, and a separate correlated motion parameter delta2 for Ag atoms to model sharpened distance distribution between it and nearest neighbor S or Ag atoms.

**Table S2**: Summary of cubic and orthorhombic models and their constrain parameters

| Model  Constrain | Pseudo cubic | Orthorhombic | Pseudo-orthorhombic (I) | Pseudo-orthorhombic (II) |
|---|---|---|---|---|
| Lattice | a=b=c | a≠b≠c | a≠b≠c | a≠b≠c |
| Symmetry | P1 | Pna21 | P1 | P1 |
| Ag | refine independently of cubic symmetry | refine following orthorhombic symmetry | refine following orthorhombic symmetry | refine independently of orthorhombic symmetry |
| SnS$_4$ | rigid body / fixed position | rigid body / fixed position | rigid body / allowed to rotate | rigid body / allowed to rotate |
| S | allowed to rotate around site | fixed position | allowed to rotate around site | allowed to rotate around site |

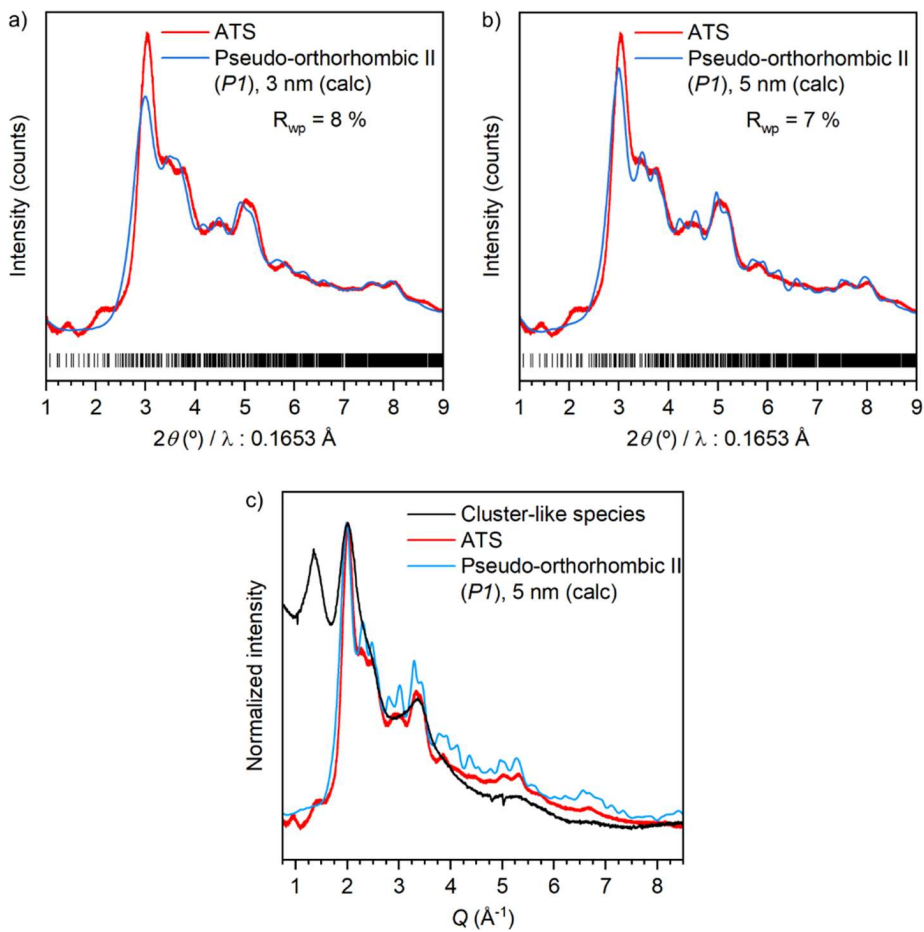

**Figure S 12**. Simulated diffraction patterns of the Pseudo-Orthorhombic (P1) structure model obtained after refinement through PDF analysis, (a) considering 3 (Rw = 8%) or (b) 5 nm (Rw = 7%) particle size, and (c) comparison with diffraction pattern of ATS and cluster-like species in Q-space.

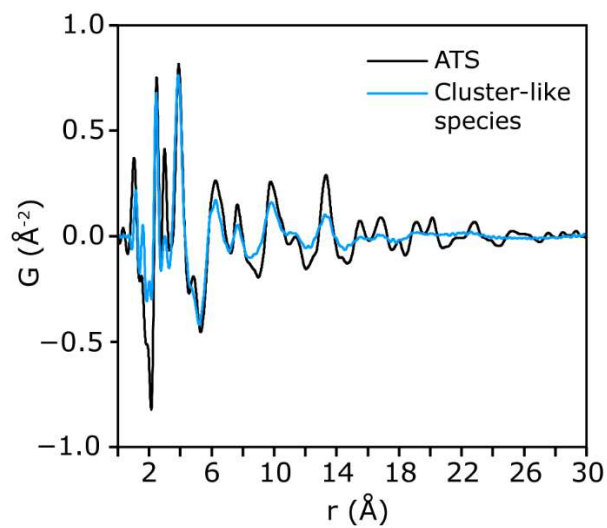

**Figure S 13**. PDF data comparing ATS to cluster-like species. There is a high similarity in the local structure (0-5Å). Thus, cluster-like species are compatible with a pseudo-orthorhombic model.

**Section 5. Physical and optical characterization for ATS@Zn NCs**

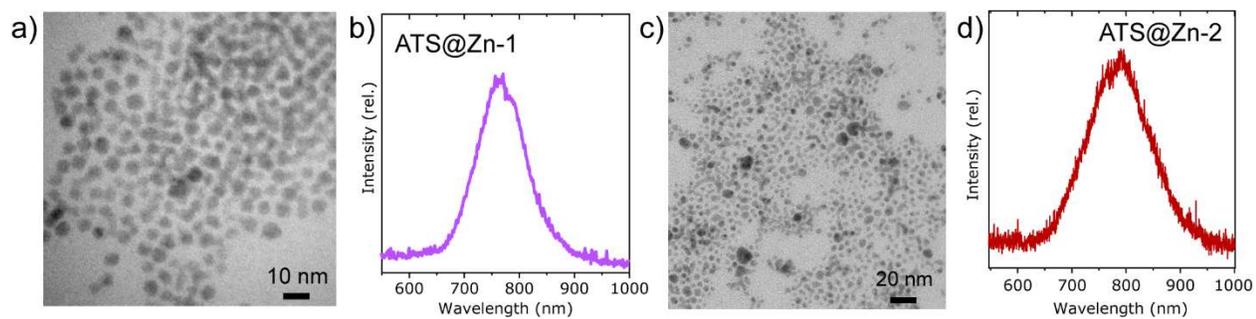

**Figure S 14**. BF TEM and photoluminescence data for ATS@Zn-1 (a and b) and ATS@Zn-2 NCs (c and d).

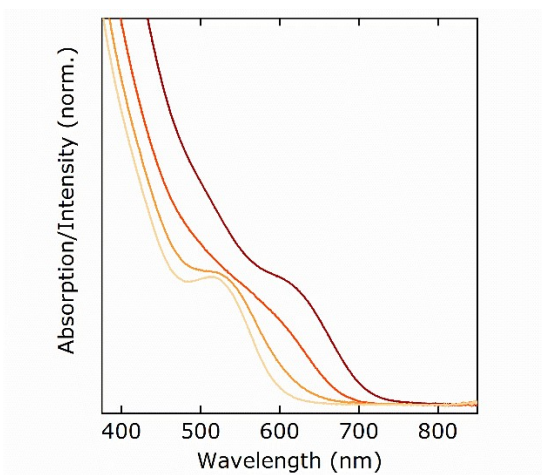

**Figure S 15**. Optical absorption spectra of aliquots taken from an ATS@Zn-1 reaction at different times after precursor injection (from 10 s to 30 min.)

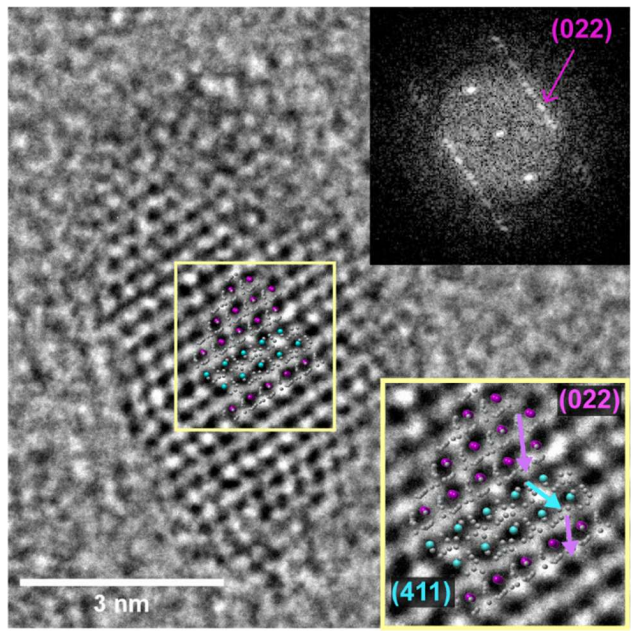

**Figure S 16**. HR-TEM image and corresponding FFT pattern for an ATS@Zn-1 NC displaying a stacking fault. Sn and Ag atoms are overlaid in pink/cyan and silver, respectively. The lattice propagates through the (022) plane with d-spacing of 0.311 nm as confirmed *via* diffraction (see inset). Then, the lattice faults and starts to propagate on the (411) plane for one lattice spacing (0.301 nm) until it returns to the (022) plane.

*Associated discussion for Figure S15*

We identified several ATS NCs with stacking faults. To get insight into the preferred planes in which the faults were originating, we calculated the d-spacing through these faults (*i.e.,* 3.11 and 3.01 Å). These values are consistent with planes (022) and (411) for orthorhombic canfieldite and constitute the major reflections in the XRD pattern, suggesting that ATS NCs might have a tendency to growth through those planes. This preferred orientation growth behaviour would be expected in such structures with anisotropic lattices. Overall, these results support our PDF analysis in confirming that our NCs have a canfieldite phase.

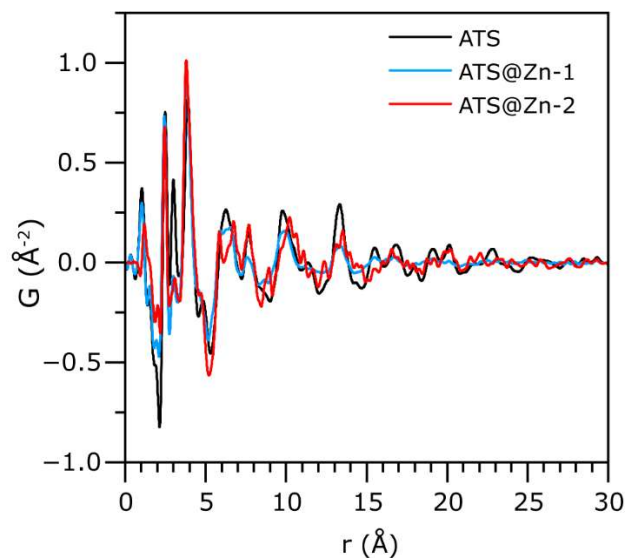

**Figure S 17**. Extended range PDF data for ATS, ATS@Zn-1, and ATS@Zn-2 showing the similarity in the local structure to a canfieldite-like phase. A subset of this data is presented as Figure 5d.

**References**

1. Yarur Villanueva, F.; Hasham, M.; Green, P. B.; Imperiale, C. J.; Rahman, S.; Burns, D. C.; Wilson, M. W. B., A Stepwise Reaction Achieves Ultrasmall $Ag_2ZnSnS_4$ Nanocrystals. *ACS Nano* **2024,** *18* (52), 35182-35201.
2. Fairley, N. F., V.; Richard-Plouet, M.; Guillot-Deudon, C.; Walton, J.; Smith, E.; Flahaut, D.; Greiner, M.; Biesinger, M.; Tougaard, S.; Morgan, D.; Baltrusaitis, J. , Systematic and Collaborative Approach to Problem Solving Using X-Ray Photoelectron Spectroscopy. *Appl. Surf. Sci.* **2021,** *5*, 100112.
3. Wieczorek, A.; Lai, H.; Pious, J.; Fu, F.; Siol, S., Resolving Oxidation States and X–site Composition of Sn Perovskites through Auger Parameter Analysis in XPS. *Advanced Materials Interfaces* **2023,** *10* (7), 2370024.
4. Villanueva, F. Y. *Understanding and Controlling the Formation Mechanism and Surface Chemistry of Lead-Free Quaternary Semiconductor Nanocrystals*. PhD thesis, University of Toronto, ProQuest, September 2024.